\documentclass{optica-article}
\journal{opticajournal} 

\articletype{Research Article}

\usepackage{lineno}
\usepackage[final]{pdfpages}

\begin{document}

\title{Wide-Field, High-Resolution Reconstruction in Computational Multi-Aperture Miniscope Using a Fourier Neural Network}

\author{Qianwan Yang\authormark{1}, Ruipeng Guo\authormark{1}, Guorong Hu\authormark{1}, Yujia Xue\authormark{1}, Yunzhe Li\authormark{1,4} and Lei Tian\authormark{1,2,3,*}}

\address{\authormark{1}Department of Electrical and Computer Engineering, Boston University, Boston, MA 02215, USA.\\
\authormark{2}Department of Biomedical Engineering, Boston University, Boston, MA 02215, USA.\\
\authormark{3}Neurophotonics Center, Boston University, Boston, MA 02215, USA.\\
\authormark{4}Department of Electrical Engineering and Computer Sciences, University of California, Berkeley, CA 94720, USA.}
\email{\authormark{*}leitian@bu.edu} 

\begin{abstract}
Traditional fluorescence microscopy is constrained by inherent trade-offs among resolution, field-of-view, and system complexity. To navigate these challenges, we introduce a simple and low-cost computational multi-aperture miniature microscope, utilizing a microlens array for single-shot wide-field, high-resolution imaging. Addressing the challenges posed by extensive view multiplexing and non-local, shift-variant aberrations in this device, we present SV-FourierNet, a novel multi-channel Fourier neural network. SV-FourierNet facilitates high-resolution image reconstruction across the entire imaging field through its learned global receptive field. We establish a close relationship between the physical spatially-varying point-spread functions and the network's learned effective receptive field. This ensures that \mbox{SV-FourierNet} has effectively encapsulated the spatially-varying aberrations in our system, and learned a physically meaningful function for image reconstruction. Training of SV-FourierNet is conducted entirely on a physics-based simulator. We showcase wide-field, high-resolution video reconstructions on colonies of freely moving \emph{C. elegans} and imaging of a mouse brain section. Our computational multi-aperture miniature microscope, augmented with SV-FourierNet, represents a major advancement in computational microscopy and may find broad applications in biomedical research and other fields requiring compact microscopy solutions.
\end{abstract}

\section{Introduction}
\label{sect:intro}

Wide-field, high-resolution imaging plays a critical role across diverse scientific disciplines. Yet, traditional microscopy, which relies on single-objective lenses, is encumbered by a fundamental trade-off between field-of-view (FOV) and resolution\cite{park2021review}. 
To circumvent these constraints, array microscopes have been developed, facilitating rapid, wide-FOV image acquisition\cite{harfouche2023imaging,son2020miniaturized,fan2019video}. These systems utilize an ensemble of microscopes, each having its own lens and image sensor, to expand the FOV without sacrificing resolution. 
Despite these advancements, the scalability and miniaturization of array microscopes are limited by their dependence on bulky, specialized image sensor arrays, which pose significant challenges in applications demanding portability and compactness, such as endoscopy\cite{fu2021future}, on-chip microscopy\cite{gorocs2012chip}, and in-vivo neural imaging~\cite{aharoni2019all}.
Recently, lensless imaging techniques have emerged as a compact alternative by employing a phase mask\cite{kuo2020chip,adams2022vivo} placed directly in front of a CMOS sensor. However, the removal of focusing optics leads to reduced contrast and signal-to-noise ratio (SNR), which restricts their sensitivity for imaging weak fluorescent signals\cite{wu2024mesoscopic,boominathan2022recent}.

\begin{figure}[t!]
\centering\includegraphics[width=\linewidth]{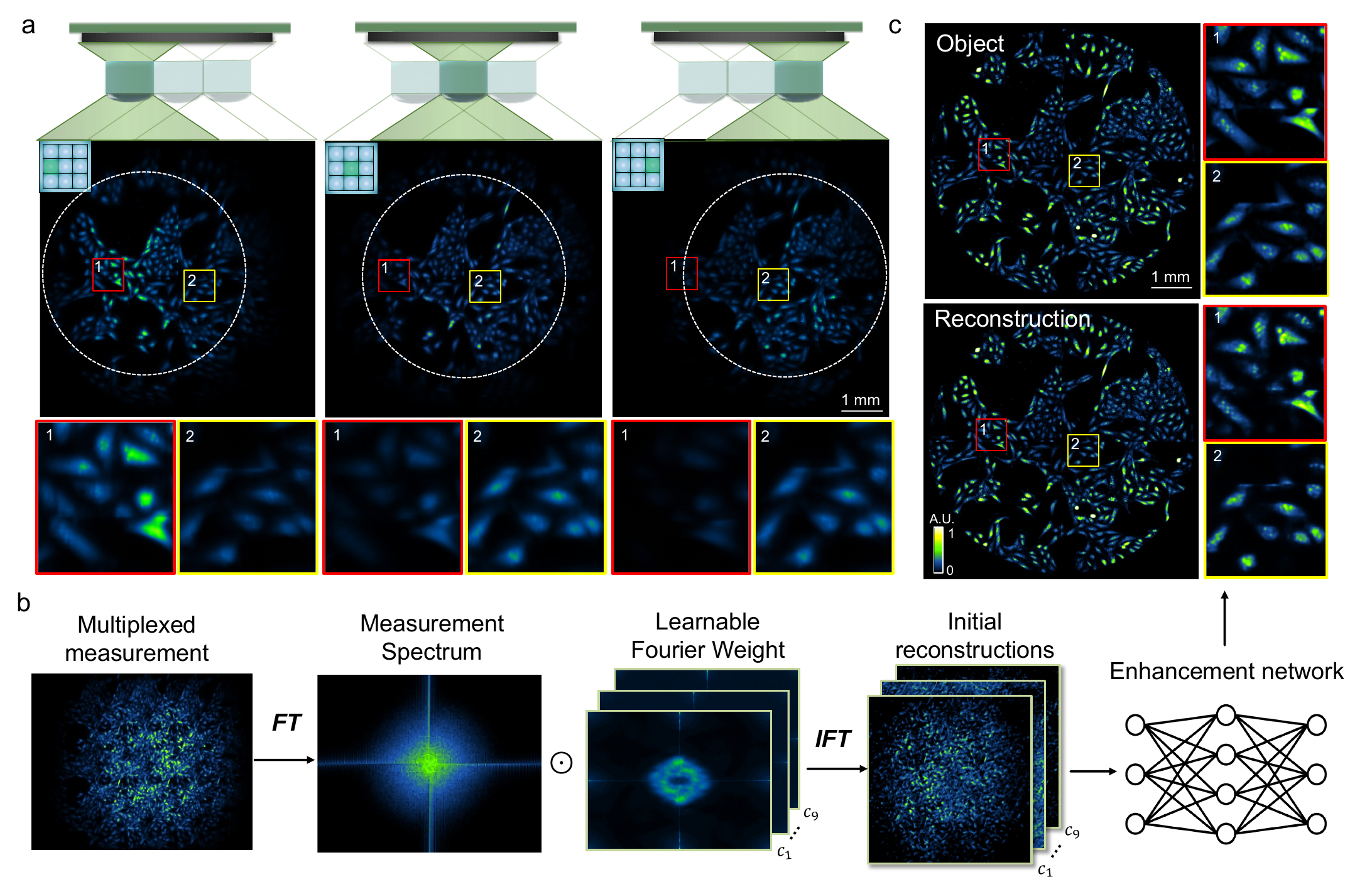}
\caption{System overview. (a) Illustrations of the capability of individual lenses within the MLA, each focusing on a specific sub-FOV, with noticeable aberrations and diminishing contrast at the  periphery. (b) The process of SV-FourierNet, where the captured image is first transformed into the Fourier domain and processed by multiple learnable filters for initial deconvolution, followed by information fusion and enhancement by a network to yield the final output. (c) The SV-FourierNet reconstruction achieves enhanced image quality and uniform resolution across an expanded FOV. 
}
\label{fig1}
\end{figure}

To address these challenges, we introduce a computational multi-aperture miniature microscope (miniscope) that delivers wide FOV and micrometer resolution on a compact and lightweight device. Drawing inspiration from  array microscopes, our design employs a  microlens array (MLA) as the sole imaging component. 
Leveraging principles from lightfield and integral imaging techniques\cite{martinez2018fundamentals}, our previous work has  demonstrated single-shot 3D imaging capability using such a design\cite{xue2020single,xue2022deep}.
This work aims to provide a new perspective of this multi-aperture imaging system in order  to address the challenges related to FOV and uneven resolution. 
Our approach is based on the premise that each lens within the MLA can capture high-quality image of a distinct, albeit limited, sub-FOV [Fig.~\ref{fig1}(a)]. While individual lenses yield images that suffer from spatially-varying aberrations and contrast loss towards the edges of their respective sub-FOVs, combining information from various microlenses enables the reconstruction of a high-resolution image over an extended FOV, as illustrated in Fig.~\ref{fig1}(a) and \ref{fig1}(c).

Departing from conventional array microscopes, our system utilizes a single image sensor to enhance device miniaturization and portability. Unlike other MLA-based multi-aperture imaging designs\cite{tanida2001thin,hu2024metalens,xu2020metalens} that employ physical barriers to isolate sub-images captured by individual microlenses, our system opts for an unobstructed arrangement to maximize the effective imaging FOV\cite{xue2020single}.
This design, however, introduces severe cross-talk between adjacent sub-images on the sensor, i.e. view multiplexing. Furthermore, the single-element MLA generates point spread functions (PSFs) that contain widely separated array foci and exhibit considerable spatial variability, i.e. non-local and spatially-varying PSFs. These complexities pose  substantial challenges to achieving high-quality image reconstruction across the extended FOV.

Our solution to simultaneously tackle view multiplexing and non-local, spatially-varying aberrations is to employ a deconvolution algorithm. 
Traditional model-based methods rely on iterative algorithms\cite{denis2015fast,debarnot2020learning,sroubek2016decomposition}, which are computationally intensive and time-consuming. Additionally, the reconstruction quality is greatly affected by manually tuned regularization terms.

Recent progress in deep-learning-based spatially-varying deconvolution methods have shown improvements in image quality and reconstruction speed\cite{yanny2022deep,wu2023real}, emerging as a promising alternative to model-based methods. 
However, deep-learning methods often suffer from two major limitations in terms of \emph{scalability} and \emph{locality bias}.
Specifically, the computational complexity and memory demands of convolutional neural networks (CNNs) increase with the input size. Thus, the direct application of CNNs to raw measurements containing tens of millions of pixels is impractical due to limited computational resources.
The common workaround involves dividing raw images into smaller patches to make CNN training manageable. 
This method, however, assumes uniformity in the data distribution, and thus \emph{invariance} in the PSFs for all image patches, overlooking the nuances due to spatially-varying aberrations. Consequently, such CNN-based methods often exhibit degraded resolution at  FOV periphery, where aberrations diverge from those in central regions\cite{xue2022deep}.

A second challenge in using CNNs for handling non-local, spatially-varying PSFs is the networks' inherent locality bias~\cite{he2016deep,szegedy2015going}, a result of the small filter sizes in convolution layers that restrict CNNs from capturing global information. Although increasing CNN's depth can broaden the ``receptive field'', this approach escalates computational costs and complicates network training. The recently proposed MultiWienerNet and MultiFlatNet~\cite{yanny2022deep,wu2023real} address these issues by initially applying deconvolution with multiple Wiener filters informed by a set of pre-calibrated, spatially-varying PSFs, followed by refinement through a CNN. Yet, this type of techniques heavily depends on the choice of these initial PSFs that need to be sufficiently representative of the underlying spatially-varying aberrations across the FOV. Our study also shows that this \emph{spatial} domain learning approach cannot efficiently reconstruct high resolution information due to the implicit bias towards low-frequency features induced by the Wiener filtering process. 
An alternative approach is the \emph{Fourier} domain Neural Network (FourierNet), which overcomes locality bias by  applying filters directly in the Fourier domain~\cite{rippel2015spectral,chen2022fourier,tian2022learned,deb2022fouriernets}. This strategy explicitly takes advantage of the \emph{global} processing capabilities of the Fourier transform, allowing FourierNet to achieve a global receptive field with minimal layers, thus efficiently sidestepping the limitations faced by traditional CNNs. While FourierNet has been successfully applied in spatially \emph{invariant} imaging systems~\cite{chen2022fourier,tian2022learned,deb2022fouriernets}, its ability to address spatially-varying systems remains unexplored. 

In this work, we introduce Spatially-Varying FourierNet (SV-FourierNet) to facilitate wide-field, high-resolution reconstruction within our computational multi-aperture miniscope. 
We show that SV-FourierNet can be applied directly on full-FOV measurements containing 12-million pixels, circumventing the uneven resolution constraints typically associated with patch-based training methods.
Furthermore, by employing multiple learnable Fourier filters [Fig.~\ref{fig1}(b)], SV-FourierNet is effective in addressing the view multiplexing and highly non-local and spatially-varying array PSFs in our multi-aperture miniscope.
The multi-channel Fourier layer in SV-FourierNet effectively performs a blind deconvolution, based on a  low-rank imaging model, to process the input measurement.
The resulting multi-channel deconvolved outputs are then fused and refined by an enhancement network, yielding the final reconstructed image. Our study shows that this \emph{Fourier} domain learning approach can more efficiently capture high resolution information as compared to the spatial domain learning methods.

To elucidate the working principle of SV-FourierNet, we employ a gradient-based method to probe the network's ``effective receptive field (ERF)''\cite{luo2016understanding}. 
This method identifies the points in the measurement that most significantly affect the prediction of a specific point on the reconstructed object, closely resembling  the PSF of our imaging system at the corresponding location. 
We show a close correspondence between SV-FourierNet's ERFs and the physical PSFs at various field positions.
This underscores that SV-FourierNet has encapsulated the spatially-varying aberrations in our system, and learned a \emph{physically meaningful} function for image reconstruction.

Compared to current existing state-of-the-art model-based and deep-learning-based  methods\cite{xue2022deep,yanny2022deep,wu2023real}, SV-FourierNet demonstrates superior reconstruction quality, consistently higher resolution across the FOV, and enhanced inference speed. Trained entirely on a physics-based simulator, we show experimentally that SV-FourierNet attains a uniform resolution of 7.8 $\mu$m across a 6.5 mm FOV. The network also provides consistent reconstruction quality across an extended depth-of-field (EDOF) over a 100 $\mu$m depth range. 
Additionally, we present wide-field, high-resolution video reconstructions of freely moving \emph{C. elegans} colonies and imaging of a weakly scattering mouse brain section.

Our computational multi-aperture miniature microscope augmented with SV-FourierNet represents a significant advancement in computational microscopy, promising wide-ranging applications in biomedical research and other areas requiring compact imaging solutions.

\section{Methods}
\subsection{The multi-aperture miniscope}
The multi-aperture miniscope is built with off-the-shelf and 3D-printed components. 
It uses an off-the-shelf plastic 3$\times$3 MLA (1 mm pitch, no. 630, Fresnel Technologies Inc.), forming a finite conjugate imaging geometry with a working distance of $\sim$7.5 mm and an MLA-sensor distance of $\sim$4.5 mm. 
Each lens within the MLA captures an image of a distinct and limited sub-FOV, while the whole array covers an extended FOV (see more details in Section 1 in Supplement 1). 
For green fluorescence imaging, a hybrid filter set is incorporated, combining an interference filter (no. 535/50, Chroma Technology) placed in front of the MLA and a long-pass absorption filter (Wratten color filter no. 12, Edmund Optics) placed after the MLA. It uses a backside-illuminated (BSI) CMOS sensor (IM226, IDS Imaging) with 12 million pixels and 1.85 $\mu$m pixel size, achieving an effective pixel size of 2.8 $\mu$m.	To validate the reconstruction results from our system, we set up a widefield  microscope to record fluorescence simultaneously with our miniscope. 
See more details in Section 9A-B in Supplement 1.

\subsection{Spatially-varying imaging model}
To bypass the need for extensive experimental data collection, we develop a highly efficient and accurate simulator to generate a comprehensive training dataset \emph{in silico}. Building upon our previous work\cite{xue2022deep}, this simulator incorporates a low-rank spatially-varying imaging model, improved with insights from a ray-tracing model, as detailed below.

\begin{figure}[t!]
\centering\includegraphics[width=\linewidth]{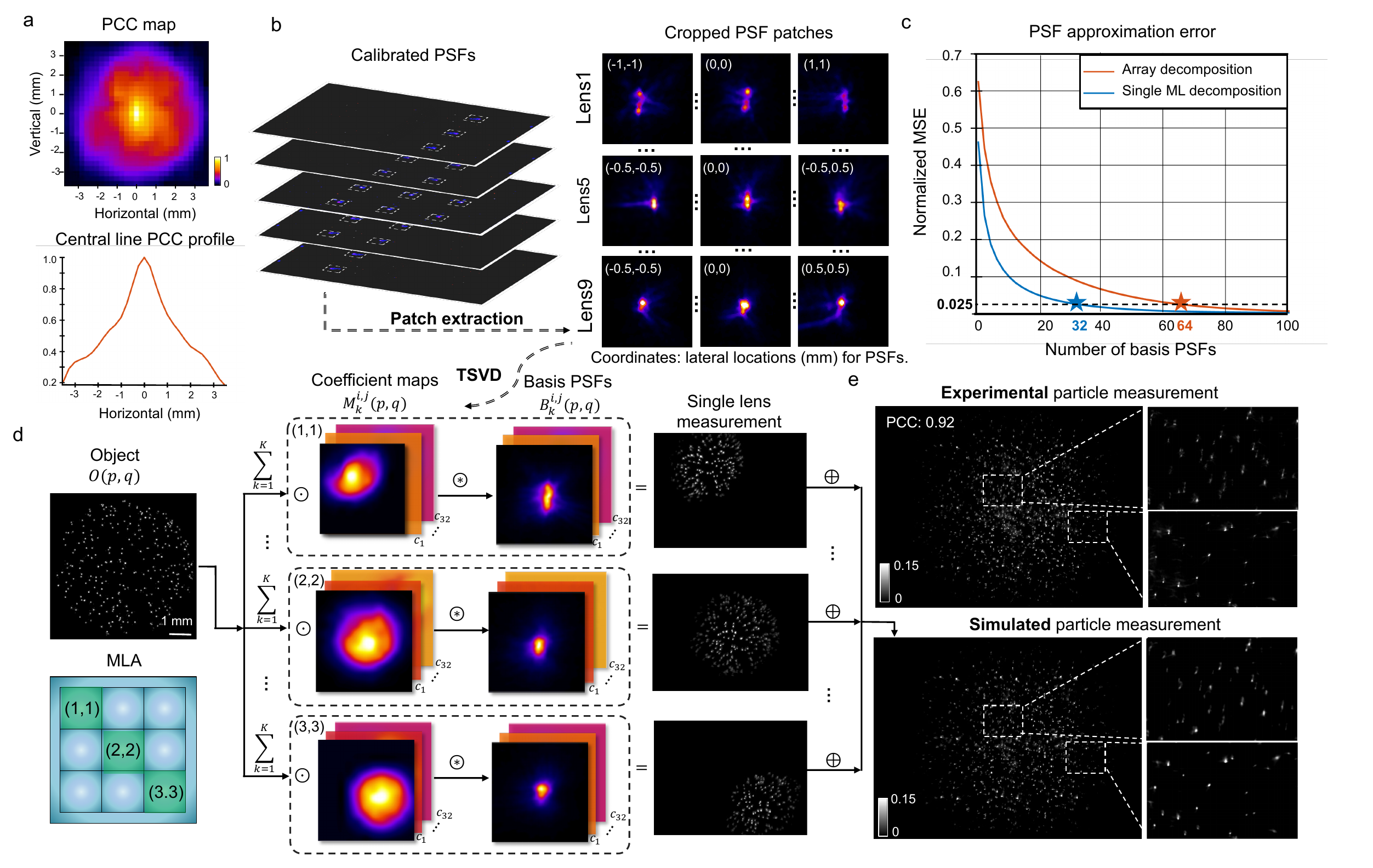}
\caption{Spatially-varying imaging model. 
(a) Spatial variance is quantified by the PCC map and the central line profile, which both exhibit a marked decline as the point source shifts away from the FOV center. (b) The low-rank  model is computed on calibrated PSFs, which are cropped around each ML's foci to enhance computational efficiency. The foci captured at different point source locations under the same ML show similar aberrations. TSVD is performed to compute the basis PSFs and coefficient maps for each ML. (c) Compared to the array decomposition method, the single-ML decomposition method reduces the model's rank by 2$\times$ while maintaining  accuracy. (d) The final imaging model involves  computing $K$ weighted convolutions between the object and the basis PSFs for each ML and a superposition of all MLs to yield the  multiplexed measurement. (e) The model is validated by comparing the simulated and physical measurements from the same object containing randomly distributed 10~$\mu$m fluorescent particles.
}
\label{fig2}
\end{figure}

We first collect a set of calibrated PSFs by scanning a single fluorescent bead across a 7 mm $\times$ 7 mm FOV with a uniform step size of 250~$\mu$m. To showcase the strong spatial variance of the PSFs, we calculate the Pearson Coefficient Correlation (PCC) map on the calibrated PSFs with respect to the on axis PSF, as shown in Fig.~\ref{fig2}(a). The PCC map exhibits a rapid radial decay due to the limited angular response of the system and spatially-varying aberrations from the MLA. The central line profile shows that the PCC decreases to 0.2 with a 3.5~mm displacement. 

Next, we develop a low-rank model in three steps. To enhance computational efficiency, we first exploit the sparsity in the array  PSF generated by the MLA and perform a cropping operation based on a ray tracing model to isolate each focal region (240 $\times$ 240 pixels) from the raw PSF measurement [Fig.~\ref{fig2}(b)]. This method relies on the calibration of the chief ray of each microlens (ML) $(i,j)$ from an on-axis point source, which establishes the ``anchor'' coordinates in both the image space $(p_{x_0}^{i,j},q_{y_0}^{i,j})$ and the object space $(x_0,y_0)$. Given the depth $z$ and the point source's lateral coordinates $(x,y)$, the corresponding PSF centroids for each ML $(i,j)$ $(p_{x}^{i,j},q_{y}^{i,j})$ is linearly proportional to the magnification $M_z$ and the object displacement:
\begin{equation}
\begin{aligned}
p_{x}^{i,j} = p_{x_0}^{i,j} - M_z\cdot(x-{x_0}),\\
q_{y}^{i,j} = q_{y_0}^{i,j} - M_z\cdot(y-{y_0}).
\label{eq1}
\end{aligned}
\end{equation}
Contrasting with cross-correlation methods typically employed to locate centroids of calibrated PSFs\cite{xue2022deep,yanny2020miniscope3d}, our ray tracing model embeds the image distortion information into the PSF patches. The distortion is characterized by the subsequent decomposition process and corrected during the reconstruction phase. 

Second, to approximate the system's spatial variance, we implement truncated singular value decomposition (TSVD) on the cropped foci for each individual ML, denoted as $H^{i,j}(p,q;x,y)$. Unlike our previous work\cite{xue2022deep} that performs a global decomposition on the array PSFs, our new method decomposes the foci of each ML independently. This is based on the observation that aberrations tend to be more consistent within a single ML but vary across MLs, largely due to the manufacturing variations, as shown in Fig.~\ref{fig2}(b). Exploiting these distinct aberrations, our single-ML decomposition method effectively reduces the rank of the imaging model and computational complexity by $2\times$ without compromising the model accuracy [Fig.~\ref{fig2}(c)]. The decomposition process can be written as 
\begin{equation}
H^{i,j}(p,q;x,y) \approx
 \sum_{k=1}^{K}M^{i,j}_k(x,y)B^{i,j}_k(p,q),
\label{eq3}
\end{equation}
where $M^{i,j}_k(x,y)$ and $B^{i,j}_k(p,q)$ denote the $k^\mathrm{th}$ coefficient map and basis PSF for ML $(i,j)$, respectively. We select $K$ = 32, achieving an approximation error of $\sim$2.5\% on the calibration set.

Finally, the spatially-varying model is implemented in two steps, as shown in Fig.~\ref{fig2}(d). First, the object is element-wise multiplied with a coefficient map and then convolved with the corresponding basis PSF for each ML. The $K$ weighted convolutions are summed to form a single-ML measurement. Next, all the single ML measurements are ``placed back'' according to the pre-calibrated anchor location $(p_{x_0}^{i,j},q_{y_0}^{i,j})$, and then summed together to construct the final multiplexed measurement. This process is described as
\begin{equation}
g(p,q) =\sum_{i,j=1}^{3}\sum_{k=1}^{K}[M^{i,j}_k(p,q)\odot O(p,q)]\circledast_{p,q}B^{i,j}_k(p,q),
\label{eq2}
\end{equation}
where $O(p,q)$ is the object's fluorescence distribution, and the image space coordinates $(p,q)$ are related to the object space coordinates $(x,y)$ by Eq. \eqref{eq1}. 

To expedite the simulation process, sub-images from individual MLs are computed in parallel, allowing the model to generate a measurement within 30 seconds. The sensor noise is sampled from a pre-calibrated mixed Poisson-Gaussian noise model. The noise is added to the measurements in the data loader as a form of ``online'' data augmentation during the training process, which ensures that the synthetic measurements closely approximate real imaging conditions (see details in Section 3B in Supplement 1).

The spatially-varying imaging model is validated by comparing the simulated and experimental measurements on the same object containing randomly distributed particles, shown in Fig.~\ref{fig2}(e) (See details in Section 3A in Supplement 1). 
The agreement between simulation and experiment, measured by PCC, achieves 0.92, indicating our model's accuracy.  Furthermore, two zoomed-in regions show that the model accurately reproduces the aberrations present in the physical measurement. 

To promote generalization of SV-FourierNet, we simulate training data by applying our low-rank model to a diverse set of biological images collected from online repositories\cite{voigt2019mesospim,todorov2020machine,hartmann2020image,srikumar2013global,dahlin2023reference}, including cells, vasculature, and brain sections, as well as synthetically generated fluorescent particles with random size, brightness, and locations, yielding around 5000 training images (see details in Section 3C in Supplement 1).

\subsection{SV-FourierNet Design}
SV-FourierNet features high scalability and effective processing of global spatial information, enabling the direct processing of raw measurements containing 12 million pixels.  This capability allows it to simultaneously address the challenges of view multiplexing and non-local, spatially-varying aberrations in our system.

\begin{figure}[ht!]
\centering\includegraphics[width=0.9\linewidth]{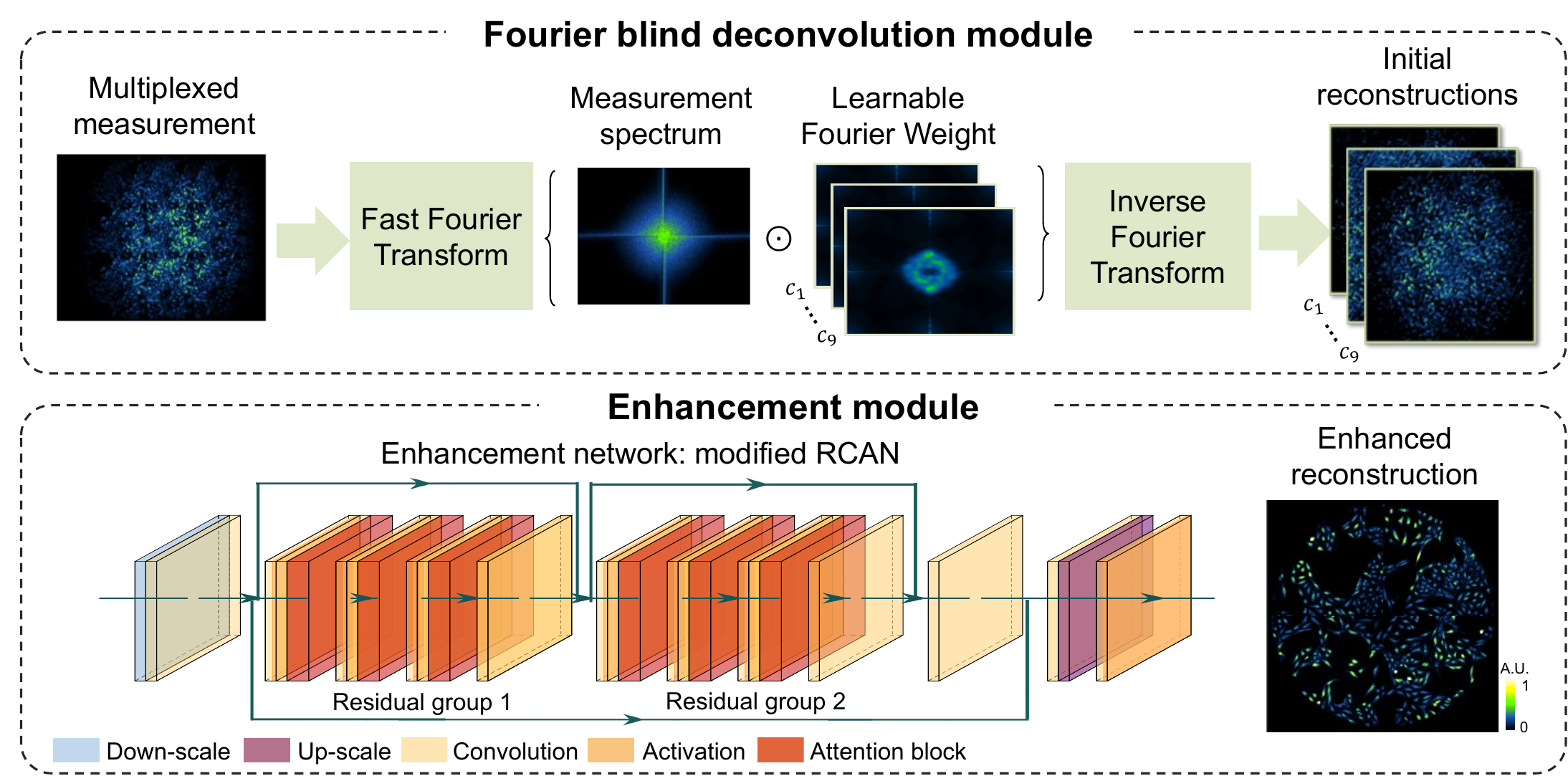}
\caption{SV-FourierNet structure. The full measurement is first sent to the Fourier blind deconvolution module to perform initial reconstructions, followed by the enhancement module to form the final reconstruction.
}
\label{fig8}
\end{figure}

SV-FourierNet comprises two core modules, including a blind Fourier deconvolution module and an enhancement module, as shown in Fig.~\ref{fig8}. The measurement is first Fourier transformed into the frequency domain, followed by element-wise multiplication with multiple learnable complex Fourier filters, whose weights are initialized with a uniform distribution $U$[0, 1]. 
The filtered results are then inverse Fourier transformed back to the spatial domain, resulting in the initial multi-channel reconstructions. 

The Fourier-domain blind deconvolution approach in SV-FourierNet, in contrast to MultiWienerNet / MultiFlatNet~\cite{yanny2022deep,wu2023real},  bypasses the need for pre-selecting PSFs and facilitates the efficient learning of a low-rank Fourier basis. 
Additionally, by leveraging the conjugate symmetry of the Fourier transform, SV-FourierNet halves the number of training parameters required compared to those needed for learning directly in the spatial domain\cite{yanny2022deep,wu2023real}. 
SV-FourierNet employs nine Fourier filters to collectively account for the spatial variance  in our system.
The number of filters is crucial as it dictates the network's capacity for nonlinearity. We have conducted an ablation study on the number of Fourier filters, which revealed that increasing the number of filters improves resolution and reconstruction quality but at a cost of memory consumption. Given the trade-off between network performance and computational cost, we have opted for nine channels to facilitate the spatially varying reconstruction (see details in Section 4C in Supplement 1).

The enhancement module employs a modified Residual Channel Attention Network (RCAN)~\cite{zhang2018image}, which includes a down-scaling module, a residual in residual (RIR) block and an up-scaling module. 
Each RIR block connects two residual groups with a long skip connection, facilitating information fusion across different scales. Channel attention modules embedded within the residual groups dynamically adjust the channel-wise feature weights, essential for synthesizing information from different Fourier channels. To prevent overfitting and to improve  generalizability, the enhancement module is designed to be lightweight, with approximately 0.18 million parameters. Additional details about the network implementation is in Section 4B in Supplement 1.  In addition, cropping and padding operations are implemented in the data loader to address the periodic ambiguity characteristic of array PSFs (see details in Section 4A in Supplement 1). 

The loss function is a sum of binary cross entropy (BCE) and Mean Squared Error (MSE):
\begin{equation}
L_\mathrm{total}=L_\mathrm{BCE} + L_\mathrm{MSE},
\label{eq4}
\end{equation}
\begin{equation}
L_\mathrm{BCE}=\frac{1}{n}\sum_{i=1}^{n}[y_{i}\log(\hat{y}_i)+(1-y_{i})\log(1-\hat{y}_i)],
\label{eq5}
\end{equation}
\begin{equation}
L_\mathrm{MSE}=\frac{1}{n}\sum_{i=1}^{n}(y_{i}-\hat{y}_i)^2,
\label{eq6}
\end{equation}
where $y_i$ and $\hat{y}_i$ represent the true and predicted values, respectively, $i$ indexes the pixel and $n$ is the total number of pixels.
This dual-loss approach leverages BCE for enhancing the reconstruction of sharp feature and MSE for capturing intensity variations, guiding the model towards accurate reconstructions in both structure and intensity. 
Gradient backpropagation from this loss updates the weights of both the Fourier filters and the enhancement module in an end-to-end manner.

We implement SV-FourierNet in PyTorch on an Nvidia A40 GPU. Training is conducted entirely on a simulated dataset with a batch size of 4. 
Optimization employs the Adam optimizer alongside a cosine annealing scheduler, with the entire training process spanning approximately 48 hours (See detailed loss curve in Section 4D in Supplement 1).

\subsection{Computation of the ERF}
To assess SV-FourierNet's ability to encapsulate non-local, spatially-varying aberrations and learn a physically meaningful inverse mapping, we calculate its ERF after the network is trained.
This involves gradient backpropagation from a pixel on the reconstructed object  to the input measurement\cite{luo2016understanding}, effectively measuring the influence of the measurement on a specific object pixel and generating a localized heatmap. 
In practice, this heatmap depends not only on the network's trained weights, but also varies with different input objects, making it input-dependent.
The ERF is the ensemble average of the heatmaps across a large dataset, effectively making it input-\emph{independent}. The SV-FourierNet's ERF is computed using a set of 520 test images (See detailed implementation in Section 5A in Supplement 1). 

We emphasize that the ERF-based method of visualizing the network's response offers more insights  into the network's learned functional mapping compared to other techniques like saliency map\cite{simonyan2013deep} and Grad-CAM\cite{selvaraju2017grad}, which often produce more ambiguous heatmaps and are designed to operate on an individual input basis.
The strength of our ERF-based analysis lies in its capacity to calculate a statistically averaged response, making the resulting heatmaps agnostic to the input. If the network fails to learn a meaningful mapping, the most likely ERF is a feature-less blur across a substantial portion of the input field, a consequence of the averaging process\cite{luo2016understanding}.
The highly localized ERFs suggests that SV-FourierNet has learned to generalize across different objects, and the learned mapping function aligns closely with the underlying imaging model. 

\subsection{Model-based reconstruction algorithms}
We derive and implement an LSV model-based algorithm based on solving a regularized least-squares problem using the alternating direction method of multiplier (ADMM) algorithm, as detailed in Section 2 in Supplement 1.

\section{Results}
\subsection{SV-FourierNet's learned ERF aligns with physical PSFs}

\begin{figure}[ht!]
\centering\includegraphics[width=\linewidth]{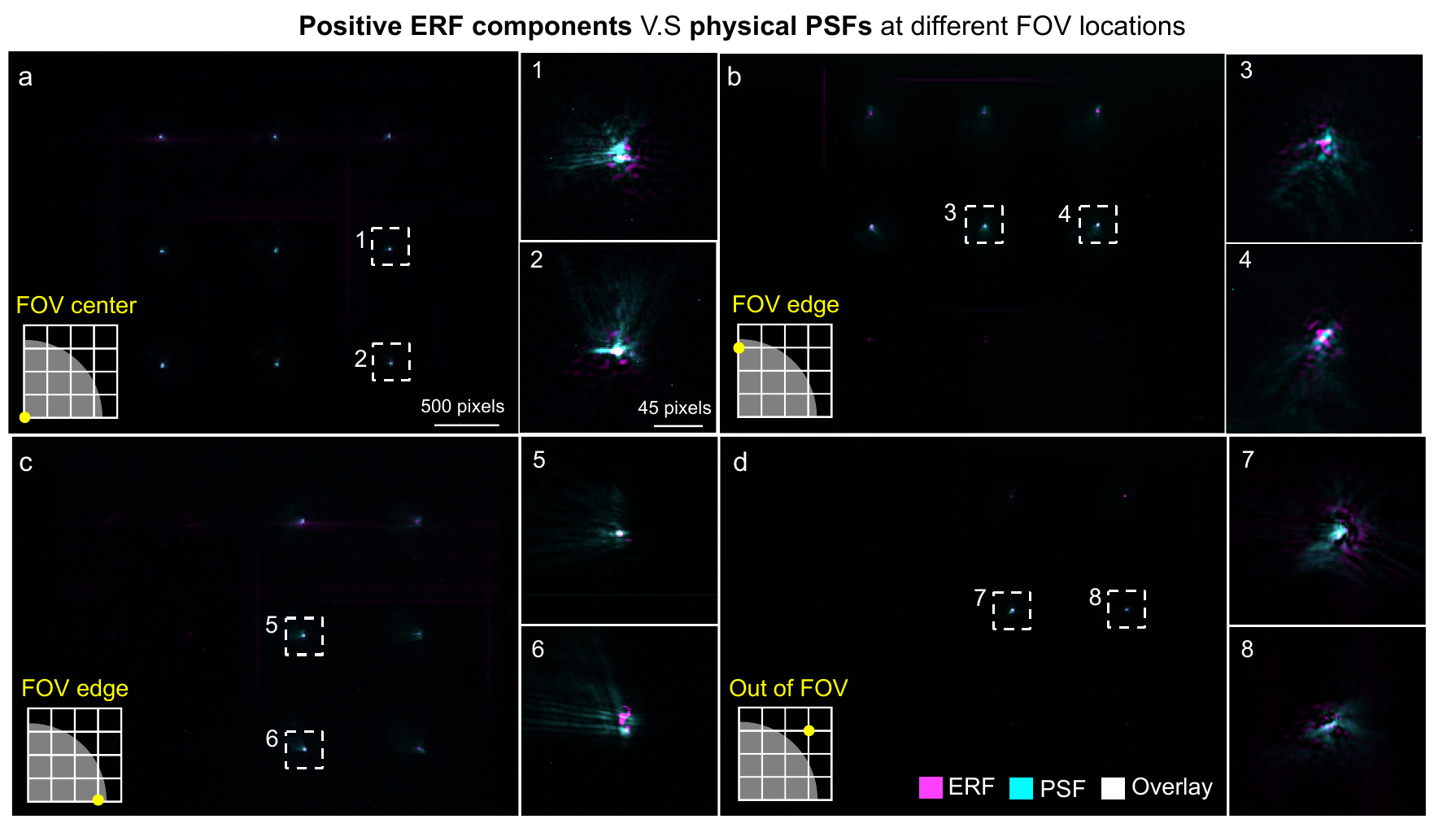}
\caption{SV-FourierNet's ERFs closely match the physically measured PSFs. The figures show the overlay between SV-FourierNet's positive ERF components (in magenta) and the physical PSFs (in cyan) at various locations: (a) the FOV center, (b-c) near the FOV edges, and (d) beyond the training FOV. For the insets, the gray region represents a quarter of the circular FOV, marked by a grid with 1 mm intervals. The yellow marker denotes the specific position of the object point. 
}
\label{fig3}
\end{figure}

We observe that the ERFs of the network closely resemble the definition of the PSFs of the physical imaging system. To discern whether SV-FourierNet has learned a meaningful reconstruction function beyond mere image feature recognition, we directly compare SV-FourierNet's ERFs with the system's PSFs across different field locations.

SV-FourierNet's ERFs contain both positive and negative components.  
The positive ERF components indicate that specific pixels in the measurement lead to an increase in intensity at the reconstructed pixel by SV-FourierNet. 
Given that PSFs in fluorescence imaging are inherently positive, it is the positive ERF components that parallel the physical PSFs' function. 
As shown in Fig.~\ref{fig3}, SV-FourierNet's ERFs are highly localized and closely match the physically measured PSFs across the field. 
At the FOV center, the ERF aligns with the physical PSF, containing a widely distributed 3 $\times$ 3 foci array, as shown in Fig.~\ref{fig3}(a). This demonstrates that SV-FourierNet has learned a global receptive field to synthesize information across the entire measurement and the network's effective response agrees with the physical response of the imaging system. 
Near the FOV edges, due to the sensor's finite size and the system's limited angular response, the physical PSFs are truncated to 2 $\times$ 3 or 3 $\times$ 2 foci array respectively, as shown in Fig.~\ref{fig3}(b) and \ref{fig3}(c). Nevertheless, the ERFs precisely localize the physical foci. 
This showcases SV-FourierNet's capability to learn the severe spatial variance in the imaging model, including handling image truncation -- a case that often induces numerical artifacts in model-based methods\cite{antipa2018diffusercam}.
In addition, when assessing a point outside the SV-FourierNet's trained FOV, the physical PSF appears as a truncated 2 $\times$ 2 foci array with noticeable contrast loss [Fig.~\ref{fig3}(d)]. However, the ERF aligns precisely with this pattern, underscoring SV-FourierNet's ability to learn a physically accurate inverse mapping. This ensures model generalization even beyond the trained FOV boundaries.
Zoomed-in views of the foci regions further elucidate the severe aberrations and spatial variability of the PSFs, along with the precise co-localization between the ERFs and the physical PSFs. A detailed quantitative analysis and the evolution of ERF throughout the training process are provided in Section 5B-C in Supplement 1.

The negative ERF components demonstrate SV-FourierNet's capability to suppress aberrations and demultiplex overlapping views in the measurement. 
These negative components, as detailed in Section 5D in Supplement 1, either surround the positive ERF to enhance resolution or create an additional periodic array foci pattern that corresponds to the MLA's periodicity.  
This pattern arises from the measurement's ambiguity caused by the MLA's periodic structure, where displacement by a distance equal to the foci array's period results in nearly identical measurements. 
The differences are only due to unique aberrations of the individual microlenses and the image truncation by the system. 
SV-FourierNet navigates this complexity by leveraging distinct aberration patterns across microlenses, accentuating the accurate object components through positive ERF components while suppressing contributions from ambiguous components via negative ERF components.

Finally, to elucidate the physical mapping function learned by the blind deconvolution module of SV-FourierNet, excluding the nonlinearity induced by the enhancement network module, we compute and visualize the equivalent spatial domain filters by transforming the learned Fourier filters to the real space, as detailed in Section 6A of Supplement 1. Notably, the visualization reveals that almost all the spatial domain filters exhibit 3 $\times$ 5 foci, which contrasts with the physical PSFs that contain 3 $\times$ 3 (or fewer) foci. To further understand this result, we compute the pseudo-inverse of the first nine basis PSFs derived via TSVD from experimentally calibrated PSFs, as detailed in Section 6B of Supplement 1. These basis filters from our physical model show a high degree of similarity to the filters learned by SV-FourierNet, each also displaying a 3 $\times$ 5 array of foci. Thus, our physical interpretation of the learned Fourier filters is a low-rank approximation of the pseudo-inverse of the LSV imaging model.

\subsection{Demonstration of consistent resolution across a wide FOV and an EDOF}
\begin{figure}[t!]
\centering\includegraphics[width=\linewidth]{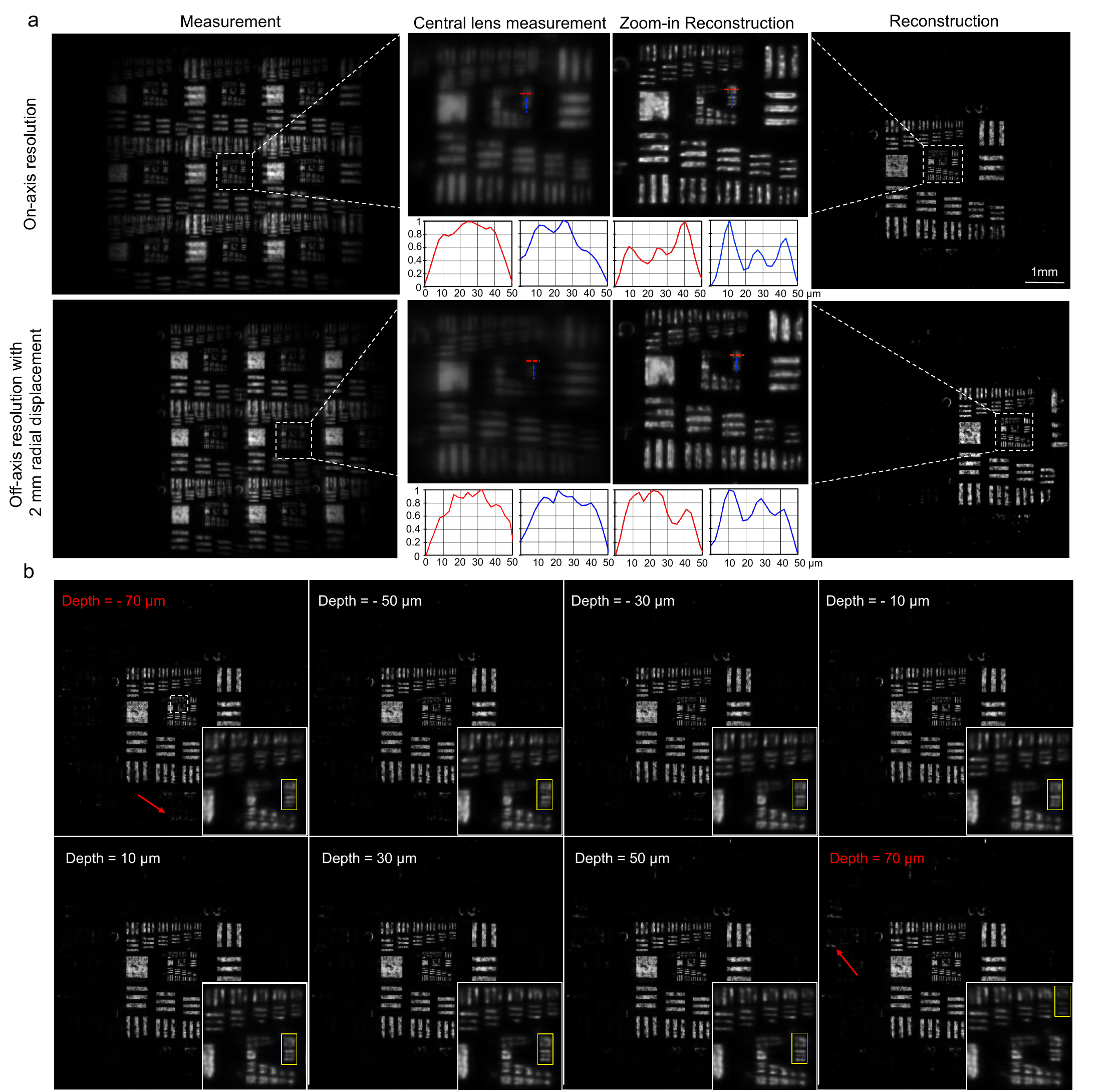}
\caption{Resolution characterization. (a) SV-FourierNet reconstruction of fluorescent resolution target  at the FOV center and edges, compared with measurements from the central lens to highlight the enhanced resolution and extended FOV achieved by our system. Line profiles confirms the system achieves a consistent 7.8 $\mu$m resolution both vertically and horizontally, at both FOV center and edges. (b) SV-FourierNet maintains a consistent resolution within a 100 $\mu$m depth range. Beyond this range, resolution begins to degrade and background artifacts appear, as indicated by the red arrow.
}
\label{fig4}
\end{figure}

To showcase our system's capability to achieve uniform resolution across a wide FOV, we characterize the resolution at both the FOV center and edges by imaging a  fluorescence resolution target, as shown in Fig.~\ref{fig4}(a). 
We also compare the reconstructions with the raw measurements from the central lens, which underscores the resolution enhancement and FOV expansion enabled by SV-FourierNet. 
In the central FOV, SV-FourierNet's reconstruction surpass those from the central lens in both resolution and contrast, illustrating the network's efficiency in effectively synthesizing information from multiple lenses for high-resolution reconstructions. 
At the FOV periphery, where the central lens exhibits pronounced aberration and vignetting artifacts, SV-FourierNet robustly reconstructs the target with high resolution, showcasing the network's capability to account for strong spatially-varying aberrations.  The line profiles further affirm SV-FourierNet's capacity to achieve a consistent 7.8 $\mu$m resolution vertically and horizontally at both central and peripheral FOV regions (See resolution characterization at additional FOV positions in Section 8C in Supplement 1). 

In addition, we characterize the reconstructed resolution of SV-FourierNet by computing the power spectral density (PSD). This analysis is performed on patches at different radial displacements and across various experimental samples. The PSD exhibits minimal variation, indicating that SV-FourierNet achieves consistent resolution across the FOV and demonstrates robustness to different experimental samples (see details in Section 8D of Supplement 1).

Furthermore, we show that the system achieves consistent resolution across an EDOF.
We characterize the system resolution across the depth by axially scanning the fluorescent resolution target with 10 $\mu$m step size, as shown in Fig.~\ref{fig4}(b). SV-FourierNet maintains consistent resolution within the [-50$\mu$m, 50$\mu$m] depth range, as indicated in the zoom-in regions labeled by white box. Beyond this depth range, the resolution begins to gradually degrade, and the background exhibits ghosting artifacts, as marked by the red arrow (See additional resolution characterization across an EDOF on FOV edges in Section 8B in Supplement 1). These results demonstrate that SV-FourierNet, although trained at a single depth, can effectively generalize to an EDOF, providing robustness against depth variations and tilt during experiments.

\subsection{SV-FourierNet achieves state-of-the-art reconstruction quality and speed}
\label{sec:benchmark}
To demonstrate the superior performance of SV-FourierNet, we compare the reconstructions from our network with the LSV-model-based ADMM algorithm and the existing state-of-the-art reconstruction networks, including CM$^2$Net\cite{xue2022deep} and MultiWienerNet\cite{yanny2022deep}. We demonstrate that SV-FourierNet achieves the best reconstruction quality with the fastest inference speed on diverse types of samples in simulation and experiments.

\begin{figure}[t!]
\centering\includegraphics[width=\linewidth]{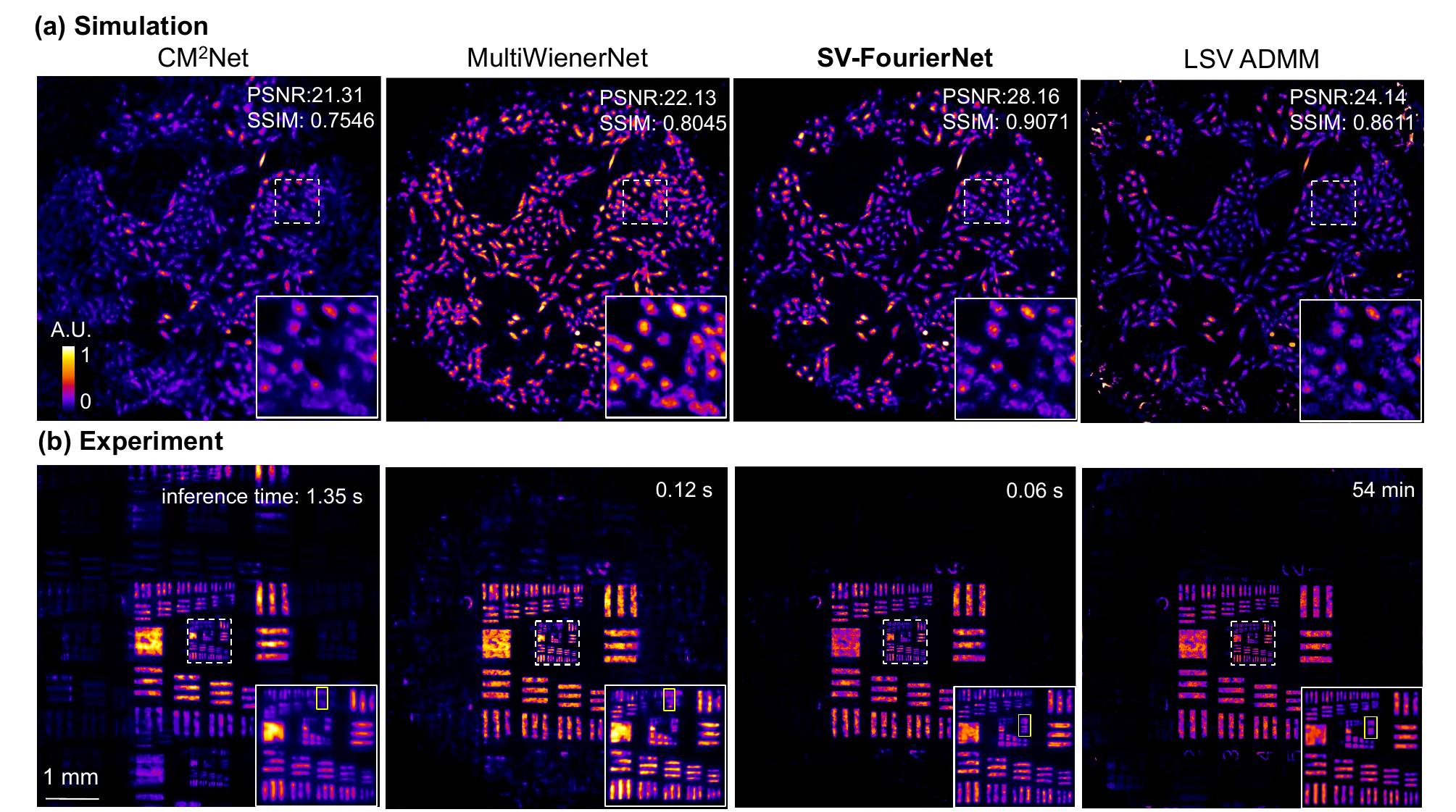}
\caption{Comparison of SV-FourierNet with existing state-of-the-art networks and LSV model-based algorithm (ADMM) on simulated and experimental datasets. (a) Reconstruction results from simulated cell measurement. SV-FourierNet achieves the best PSNR and SSIM. (b) Reconstruction result from experimental measurement on a resolution target. SV-FourierNet provides the best spatial resolution (highest spatial resolution achieved by each method is marked by the yellow box) and the fastest inference speed.
}
\label{fig5}
\end{figure}

To ensure a fair comparison between the deep learning models, we employ the same loss function and training strategy across all models. Each model is first trained on the same dataset for 48 hours. The optimal models achieving the highest SSIM are then chosen for the benchmark comparison (see details in Section 7 in Supplement 1). Moreover, the architecture of the enhancement network within MultiWienerNet is modified to be the same as that of SV-FourierNet. The detailed architecture of the benchmark networks can be found in Section 7 in Supplement~1.

We compare the reconstruction results from CM$^2$Net, MultiWienerNet, SV-FourierNet and LSV ADMM algorithm on both simulated testing data and experimental measurements, shown in Fig.~\ref{fig5}. From metrics evaluation, SV-FourierNet achieves the best PSNR and image quality (additional quantitative analysis on simulated testing data is provided in Section 7 in Supplement~1). 
From the resolution target reconstruction results, SV-FourierNet demonstrates superior resolution compared to CM$^2$Net and MultiWienerNet and is comparable to the LSV ADMM algorithm. Notably, at the edge of the FOV, SV-FourierNet maintains consistently high resolution, whereas other techniques exhibit degraded resolution, as detailed in Section 8A of Supplement 1.
By visual inspection, CM$^2$Net suffers from severe ghosting artifacts and a degradation in performance at the FOV's peripheral regions. This is attributed to CM$^2$Net's patch-based training strategy, which neglects the strong shift variance between patches. 
Similar artifacts are observed in reconstructions from model-based deconvolution algorithm using a spatially invariant model (see details in Section 2A in Supplement 1), demonstrating the importance of accounting for spatially-varying degradation in the reconstruction process. 

MultiWienerNet improves reconstruction performance at the FOV edge through multiple initial Wiener deconvolutions. However, it fails to preserve high-resolution details and suffers from pronounced background artifacts compared to SV-FourierNet.
In addition to  superior image quality, SV-FourierNet outperforms other methods with its enhanced reconstruction speed and less computational load. By leveraging the fast Fourier transform's computational efficiency, SV-FourierNet achieves a 20$\times$ speedup compared to CM$^2$Net. Furthermore, by utilizing the Fourier transform's symmetry property for real-valued signals and avoiding the additional computational costs associated with Wiener deconvolution, SV-FourierNet reduces 2$\times$ memory consumption and achieves a 2$\times$ speedup over MultiWienerNet (See additional comparisons for experimental and simulation results in Section 7 in Supplement 1).

To understand the superior performance of SV-FourierNet and our training strategy, we conducted extensive ablation studies, leading to two main insights: the impact of PSF/filter initialization and the choice of learning domain (spatial vs Fourier). 
First, initializing with physical PSFs, regardless of the learning domain, results in reduced resolution and pronounced noisy artifacts (see Section 6D in Supplement 1). This is because physical PSF initialization introduces unwanted locality biases and inherent noise in the PSF measurements. Inspecting the learned filters in MultiWienerNet initialized with physical PSFs (Section 6C in Supplement 1) reveals that the learning process primarily adds non-interpretable noisy features to the filters, similar to those reported in \cite{yanny2022deep}. This argument is further corroborated by the ERF of MultiWienerNet (Section 5B in Supplement 1), which shows severe noisy artifacts in the learned functions. Similar artifacts are observed when SV-FourierNet is trained with the same physical PSF initialization, instead of the random initialization used in our method. Replacing the physical PSF initialization with the same random initialization allows MultiWienerNet to converge to more physically meaningful filters, similar to those learned by SV-FourierNet, but with reduced resolution.
These results lead to our second insight: learning in the spatial domain generally results in inferior resolution compared to learning in the Fourier domain. This can be explained by two factors: the smooth PSF spectrum improves the conditioning of the learning problem, and Fourier domain learning eliminates an unnecessary Fourier transform operation in the spatial domain approach, reducing noise and error propagation.

\begin{figure}[h!]
\centering\includegraphics[width=\linewidth]{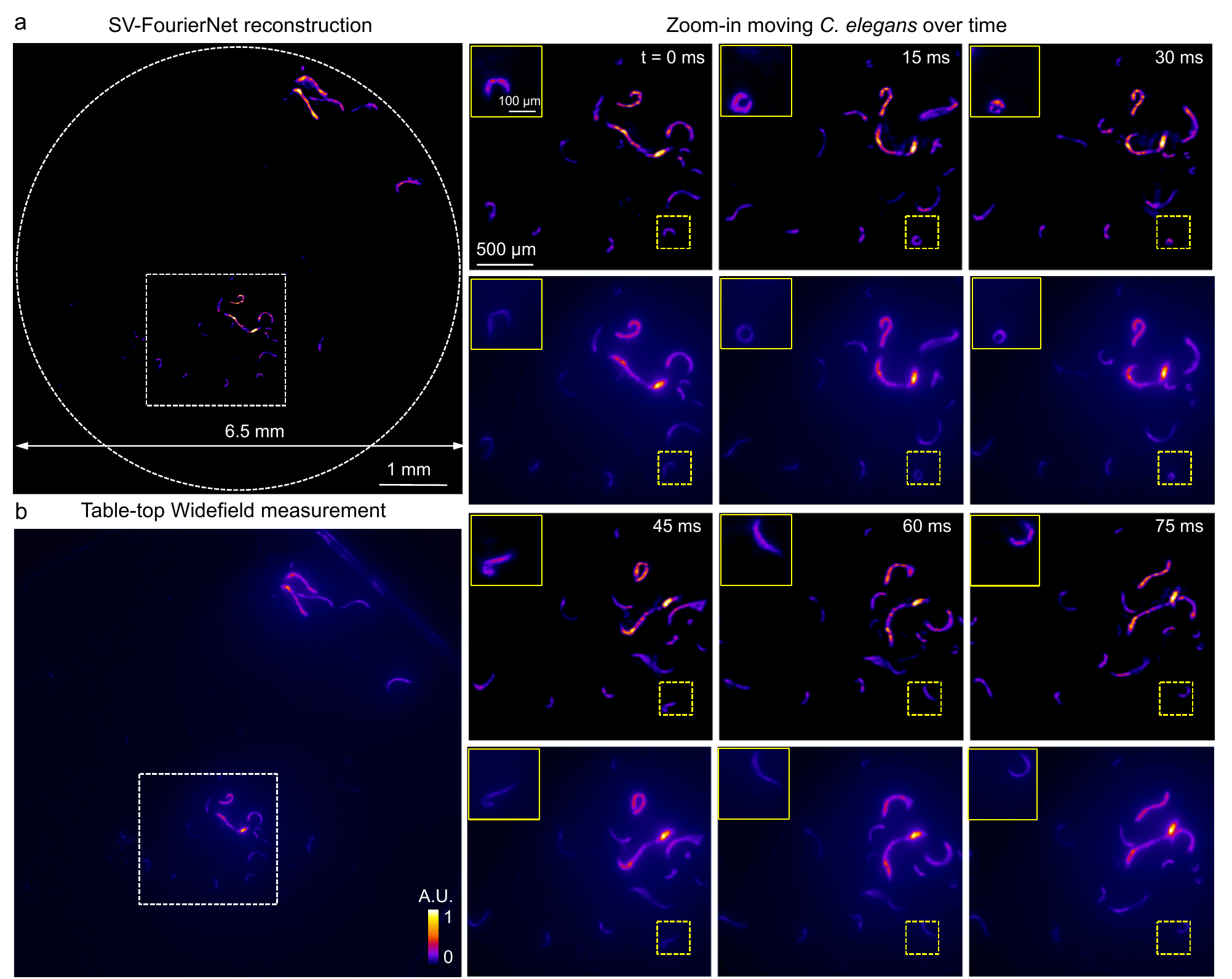}
\caption{Imaging of a freely-moving \emph{C. elegans} colony. (a) SV-FourierNet's reconstruction captures the rapid movement of the \emph{C. elegans} across a 6.5 mm FOV. (b) The reconstructions agree well with the concurrent widefield measurements across the entire FOV and zoomed-in regions, demonstrating our system’s capacity to robustly image complex biological dynamics on a large scale.
}
\label{fig6}
\end{figure}

\subsection{Experimental demonstration on a colony of freely moving C. elegans}

To showcase our system's capability in imaging large-scale dynamic biological processes, we image colonies of freely moving \emph{C. elegans}. Details about the sample preparation are provided in Section 9D in Supplement 1.
The SV-FourierNet reconstruction [Fig.~\ref{fig6}(a)] and concurrent widefield measurements [Fig.~\ref{fig6}(b)] are compared.

We compare the full-FOV reconstruction of a single frame from SV-FourierNet with the corresponding widefield measurement. The reconstruction accurately recovers all the \emph{C. elegans}, including those in central and edge FOV regions, showcasing consistent performance across the extensive FOV. 
To further demonstrate our system is robust to dynamic movements and varying local contrast, we compare sequences of SV-FourierNet reconstructions and widefield measurements for the area marked by the white box. 
These comparisons demonstrate that SV-FourierNet is capable of accurately recovering not only the larger, high-contrast \emph{C. elegans} but also the younger specimens with faint fluorescence signals, as indicated by the yellow dashed box. 
Additionally, our system effectively recovers the dynamic movements of the \emph{C. elegans} population, with a video shown in Visualization 1.

\begin{figure}[t!]
\centering\includegraphics[width=\linewidth]{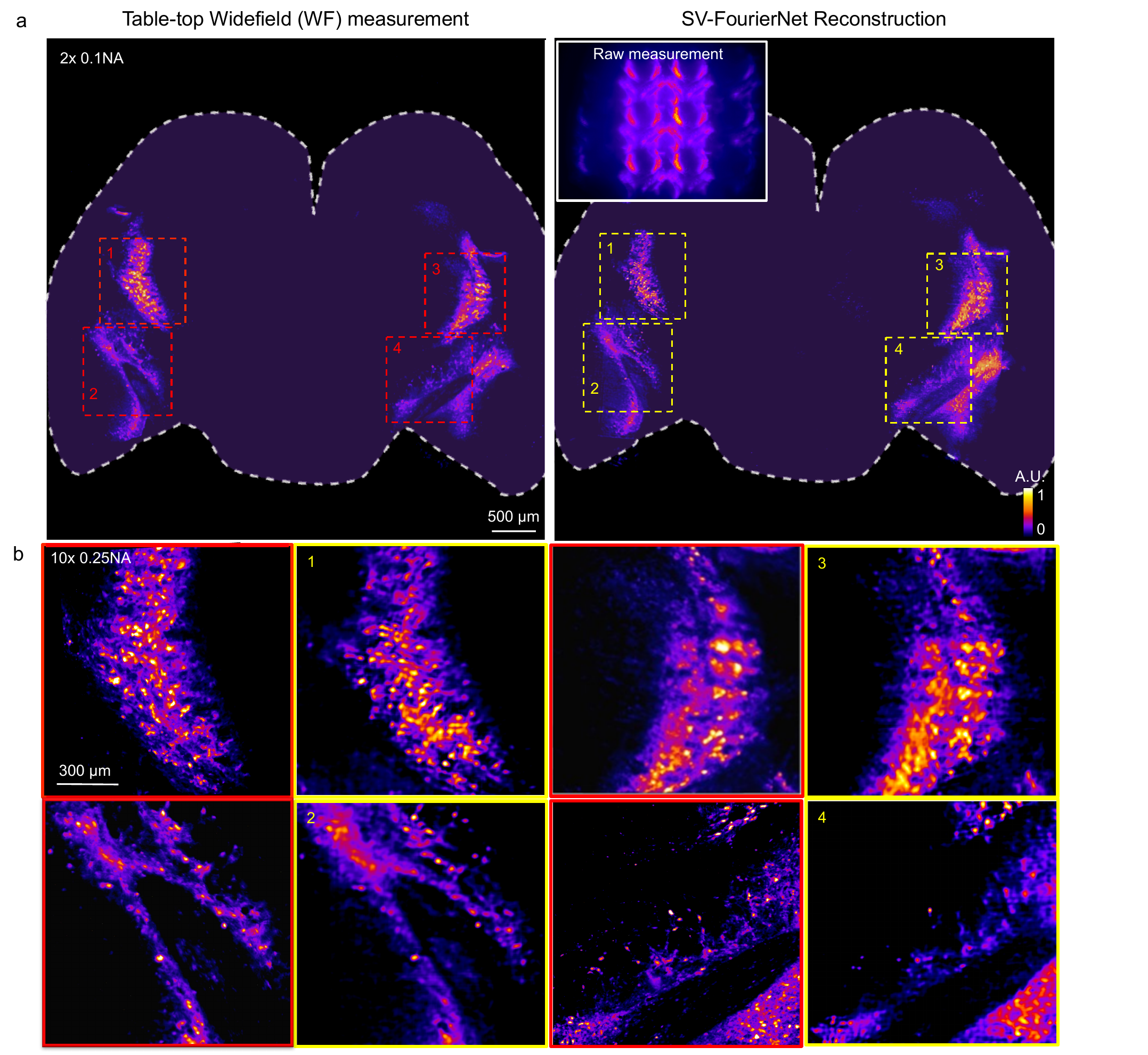}
\caption{Imaging of a 75 $\mu$m thick brain section. (a) SV-FourierNet reconstruction (with the raw measurement as an inset) aligns with the widefield measurement from a 2$\times$ 0.1NA objective, showcasing the network's ability for accurate reconstruction across an entire 6.5 mm brain section. (b) In the zoomed-in regions, SV-FourierNet reconstructions (with yellow image borders) agree with widefield measurements from a 10$\times$ 0.25NA objective (with red image borders), demonstrating our network's capacity to achieve reconstruction at cellular resolution.
}
\label{fig7}
\end{figure}

\subsection{Experimental demonstration on fixed brain section}

To demonstrate our system's capability for high-quality imaging of complex biological tissues, we image a 75 $\mu$m thick, weakly scattering brain section containing GFP-expressing neurons in the bed nucleus of the stria terminalis (BNST). The BNST, crucial in regulating emotional and stress responses, is bilaterally located on either side of the brain, approximately 2.5 mm apart, with each region extending about 2 mm. These regions are characterized by their intricate spatial structure and varying neuron density levels (detailed sample information is in Section 9C in Supplement 1).

First, we evaluate the full-FOV SV-FourierNet reconstruction by comparing it with a 2$\times$ widefield measurement. Upon visual inspection, SV-FourierNet reliably reconstructs the bilateral structure of the labeled brain regions, as shown in Fig.~\ref{fig7}(a). We then compare the reconstructions of four zoomed-in regions from both sides of the brain section with 10$\times$ widefield measurements (shown in Fig.~\ref{fig7}(b)). SV-FourierNet accurately reconstructs neuronal structures from both densely and sparsely populated areas (An additional overlay comparison between SV-FourierNet reconstruction and 2$\times$ widefield measurement is provided in Section 9E of Supplement 1). 
These results demonstrate that SV-FourierNet, trained entirely on simulated data,  can generalize to complex brain tissues, provide cellular-resolutions across the entire brain section, and is robust against weak tissue scattering.

\section{Conclusion}
We present a computational multi-aperture miniscope augmented with SV-FourierNet that performs single-shot high-resolution wide-FOV imaging. 
Our system is simple and compact, using only a single MLA for imaging. 
Unlike studies focusing on exploiting the 3D imaging capability with an MLA\cite{xue2020single}, we introduce a novel computational framework to  expand the FOV while ensuring uniform high resolution throughout. 
Our main innovation is SV-FourierNet, which employs multiple learnable Fourier filters to synthesize information over an extended FOV, addressing spatially-varying aberrations and view multiplexing inherent to our system. 
Furthermore, we demonstrate a novel utility of the network's ERF in elucidating the mapping function it learns.
We show that SV-FourierNet learns a physically meaningful reconstruction function for the spatially-varying imaging model, achieving uniform resolution throughout the extended FOV. 
Trained exclusively on simulated data, SV-FourierNet exhibits robust generalization  to diverse experimental samples, as demonstrated on resolution targets, fluorescent beads, live \emph{C. elegans}, and brain tissue sections.
This underscores SV-FourierNet's potential for a wide array of applications within biomedical research and beyond.

We show SV-FourierNet's superior capability to demultiplex cross-talks between regular MLAs and tackle severe spatially-varying degradation in 2D imaging scenarios. 
This  principle also holds promise for expanding the FOV in 3D imaging applications. 
For example, the Fourier lightfield microscope employs a field stop to prevent view multiplexing between lenses, which inevitably reduces the FOV in proportion to the lens count, thus limiting its imaging capacity for large-scale biological processes\cite{hua2021high}. 
Our strategy suggests the possibility of removing the field stop and harnessing view multiplexing to increase the FOV without compromising spatial resolution. 
In 3D multi-aperture microscopes with view multiplexing, like CM$^2$, existing 3D reconstruction algorithms suffer from resolution degradation near the FOV edges\cite{xue2022deep}.
Extending SV-FourierNet to 3D imaging may substantially improve the imaging volume with uniform 3D resolution.
Thus, we anticipate that SV-FourierNet could offer a novel paradigm for single-shot wide-FOV high-resolution volumetric imaging.

SV-FourierNet effectively learns global shift-variant inverse filters in the frequency domain, which not only significantly enhances image quality with higher spatial resolution but also substantially reduces the computational burden compared to spatial domain learning.
This reduction is crucial for model scalability, especially when handling large volumes of data. To further decrease model size and computational load, a promising direction involves exploring the sparse nature of the system's PSFs. Such reduction in the model size will enable SV-FourierNet to be adapted to more challenging applications, such as volumetric shift-variant deconvolution.

For the first time to our knowledge, we establish a connection between the ERF of a deconvolution network and the underlying physical imaging model.
This enables a straightforward visualization of the function learned by the network, alleviating the ``black box'' aspect often associated with the deep learning networks.
Furthermore, the ERF can offer a promising tool to inform network design, allow for assessing network performance and alignment between the trained deconvolution model and the physical optical system by visualizing deviations of the ERFs from the physical PSFs. 

Our accurate and efficient physics simulator eliminates the need for extensive and laborious physical data collection, facilitating a broad data distribution, which is particular impactful when ground truth is difficult to obtain. These simulated training data are crucial for the network's ability to generalize robustly across different experimental data types. To broaden the platform's application to more complex scattering biological tissues, it is possible to integrate the scattering process into the imaging model\cite{alido2024robust}. With training data derived from this enhanced imaging model, we anticipate the platform's utility to more challenging biomedical imaging applications.

\bigskip
\begin{backmatter}

\bmsection{Funding}
This project has been made possible in part by National Institutes of Health (R01NS126596) and a grant from 5022 - Chan Zuckerberg Initiative DAF, an advised fund of Silicon Valley Community Foundation.

\bmsection{Acknowledgments} 
The authors thank Boston University Shared Computing Cluster for proving the computational resources, Ian G. Davison, Kevin J Monk, and Brett T DiBenedictis for providing the brain slices, as well as Christopher A Gabel and Andrew S. Chang for providing the \emph{C. elegans} sample. We also thank BU CISL group members for insightful discussions and Adelina Chau and Jamin Xie for their assistance with benchmark network tuning.

\bmsection{Disclosures}
The authors declare no conflicts of interest.

\bmsection{Data Availability}
The SV-FourierNet implementation, and the pretrained model are available at: \href{https://github.com/bu-cisl/SV-FourierNet}{https://github.com/bu-cisl/SV-FourierNet}.

\bmsection{Supplemental document}
See Supplement 1 for supporting content.

\end{backmatter}
\bibliography{reference}



\section*{Supplementary material}

## 1. COMPARISON OF SINGLE LENS AND MICROLENS ARRAY IMAGING

This section aims to provide an intuitive understanding of how computational multi-aperture miniscope platform contributes to an extended FOV. Initially, we simulate a central lens measurement of an object containing cell feature structures with the shift variant forward model introduced in the main text (shown in Fig. S1(a)). Upon full FOV inspection, due to light vignetting and off-axis aberrations, the imaging area for a single lens is constrained to less than 5 mm area. From the zoomed in region, the single lens measurement at the center FOV can reveal small features within each cell. However, the image quality significantly degrades near the FOV edges due to aberrations.

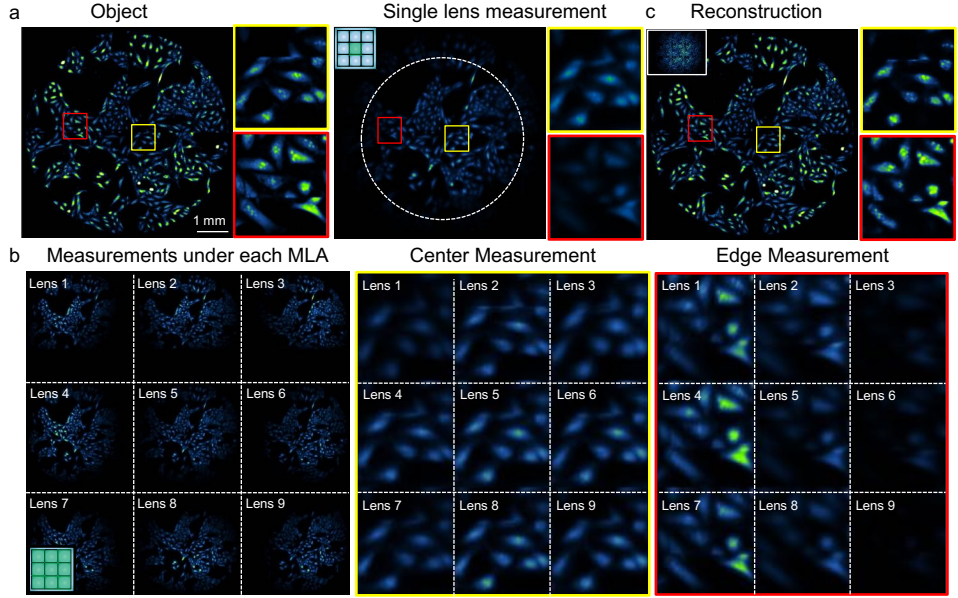

**Fig. S1. Comparison of single lens measurement, MLA measurement and SV-FourierNet reconstruction**. a: Comparison of ground truth and single lens measurement. The FOV of a single lens is limited to approximately 5mm due to light vignetting. In contrast with the center FOV measurement, the image quality at the FOV edge for the single lens is significantly degraded due to off-axis aberrations. b: The montage of measurements across different lenses. Each lens in the MLA focuses at a different subsection of the FOV, ensuring both the central and edge FOV region are captured with high image quality. c: The SV-FourierNet synthesizes the high quality measurement from different lenses, successfully recovering information across 6.5 mm FOV with improved contrast and resolution as compared to single lens measurement.

The computational multi-aperture miniscope overcomes this limitation by utilizing a microlens array (MLA). Fig. S1(b) shows the montage of measurements under different lenses. Each lens within this array captures a clear portion of the FOV, ensuring the peripheral information, neglected by the central lens shown in Fig. S1(a), is accurately recorded by lens 1,4 and 7 at high resolution. Consequently, the computational multi-aperture miniscope gathers high-resolution data across an expansive FOV in a multiplexed manner. Building on this, the SV-FourierNet effectively synthesizes the information from these measurements and provides an aberration-free high-resolution wide-FOV reconstruction, as shown in Fig. S1(c).

## 2. ADDITIONAL DETAILS ON THE IMPLEMENTATION OF THE SHIFT-VARIANT MODEL-BASED RECONSTRUCTION

### A. Comparison of shift-invariant and shift-variant model-based reconstruction

To demonstrate the importance of accounting for shift variance in achieving high-quality large FOV reconstructions, we compared a linear shift-invariant (LSI) model-based algorithm[1] with a linear shift-variant (LSV) approach. Drawing from the shift-variant forward model described in the manuscript, we have reformulated it into a compact matrix form:

$$y = BCx, \tag{S1}$$

where B is the convolution matrix that represents the basis PSFs and C is the corresponding coefficient volume at the reconstruction grid. The inverse problem can be written as an optimization problem consisting of data fidelity term and regularization term:

$$\hat{x} = \underset{x}{argmin} \frac{1}{2}\|BCx - y\|_2^2 + \mathbb{I}_+\{x\} + \tau_1\|x\|_1^1 + \tau_2\|\Psi x\|_1^1. \tag{S2}$$

The regularization consists of three commonly used terms: non-negativity, $l_1$ norm and total variation. These three terms individually enforces positive, sparsity and smoothness constraint on the reconstruction. To solve this regularized least-squares problem efficiently, we adopt the alternating direction method of multiplier (ADMM) algorithm by splitting the problem into a sequence of sub optimizations, where each sub optimization either has a closed-form solution or a fast proximal operator.

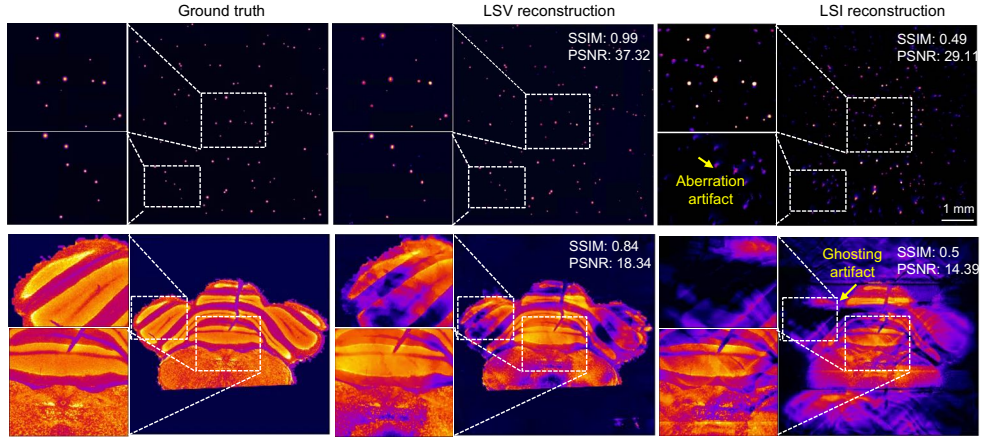

**Fig. S2. Comparison of LSI and LSV model based reconstruction**. Ground truth, LSV and LSI reconstruction results for sparse particles and brain slice. Reconstruction from both center and edge FOV are zoomed in for comparison. Notably, the LSV reconstruction demonstrates significantly enhanced image quality, particularly at the FOV edge, addressing the aberration and ghosting artifacts that are existed in LSI reconstructions.

We test the LSV model based reconstructions on both simulated sparse particles and dense brain slice measurement. And they are compared with the results obtained from previous developed LSI ADMM reconstruction algorithm[1]. For a fair comparison, both the regularization strength, ADMM initialization, step size and maximum iterations are consistent across both methods. From full FOV inspection and metric evaluation, the LSV reconstruction outperformed the LSI approach for both sparse particles and dense brain slice. The zoomed in regions show that while LSI and LSV reconstruction yield comparable reconstruction at the FOV center, the LSV reconstruction notably excels at the FOV edges, addressing the ghosting and strong aberration artifacts that are evident in the LSI reconstruction.

### B. Detailed implementation on shift-variant ADMM algorithm

Given the inverse problem stated in Eq. (S2), we can re-write the optimization problem into augmented Lagrangian form:

$$\underset{x,v,u,w,z}{argmin} \frac{1}{2}\|Bv - y\|_2^2 + + \mathbb{I}_+\{u\} + \tau_1\|w\|_1^1 + \tau_2\|\Psi z\|_1^1+$$

4

$$\frac{\mu_1}{2}\|Cx - v + \frac{\gamma_1}{\mu_1}\|_2^2 + \frac{\mu_2}{2}\|x - u + \frac{\gamma_2}{\mu_2}\|_2^2 + \frac{\mu_3}{2}\|x - w + \frac{\gamma_3}{\mu_3}\|_2^2 + \frac{\mu_4}{2}\|\Psi x - z + \frac{\gamma_4}{\mu_4}\|_2^2 \quad (S3)$$

**Table S1.** Primals, duals and parameters

|  | data fidelity | non-negativity | total variation | l1 |
|---|---|---|---|---|
| Primal: | $v = Cx$ | $u = x$ | $w = x$ | $z = \Psi x$ |
| Dual: | $\gamma_1$ | $\gamma_2$ | $\gamma_3$ | $\gamma_4$ |
| Slack: | $\mu_1$ | $\mu_2$ | $\mu_3$ | $\mu_4$ |

### B.1. Update primals

$$v^{t+1} = \underset{v}{\operatorname{argmin}} \frac{1}{2}\|Bv - y\|_2^2 + \frac{\mu_1}{2}\|Cx^t - v + \frac{\gamma_1^t}{\mu_1}\|_2^2 \quad (S4)$$

$$v^{t+1} = \left(B^T B + \mu_1 I\right)^{-1} \left(B^T y + \mu_1 Cx^t + \gamma_1^t\right) \quad (S5)$$

$$u^{t+1} = \underset{u}{\operatorname{argmin}} \mathbb{I}_+\{u\} + \frac{\mu_1}{2}\|x^t - u + \frac{\gamma_1^t}{\mu_1}\|_2^2 \quad (S6)$$

$$u^{t+1} = max\{0, x^t + \frac{\gamma_1^t}{\mu_1}\} \quad (S7)$$

$$w^{t+1} = \underset{w}{\operatorname{argmin}} \tau_1 \|w\|_1^1 + \frac{\mu_2}{2}\|x^t - w + \frac{\gamma_2^t}{\mu_2}\|_2^2 \quad (S8)$$

$$w^{t+1} = Ts\{x^t + \frac{\gamma_2^t}{\mu_2}, \frac{\tau_1}{\mu_2}\} \quad (S9)$$

$$z^{t+1} = \underset{z}{\operatorname{argmin}} \tau_2 \|z\|_1^1 + \frac{\mu_4}{2}\|\Psi x^t - z + \frac{\gamma_4^t}{\mu_4}\|_2^2 \quad (S10)$$

$$z^{t+1} = Ts\{\Psi x^t + \frac{\gamma_4^t}{\mu_4}, \frac{\tau_2}{\mu_4}\} \quad (S11)$$

$$x^{t+1} = \underset{x}{\operatorname{argmin}} \frac{\mu_1}{2}\|x - C^+(v^{t+1} - \frac{\gamma_1^t}{\mu_1})\|_2^2 + \frac{\mu_2}{2}\|x - u^{t+1} + \frac{\gamma_2^t}{\mu_2}\|_2^2 +$$
$$\frac{\mu_3}{2}\|x - w^{t+1} + \frac{\gamma_3^t}{\mu_3}\|_2^2 + \frac{\mu_4}{2}\|\Psi x - z^{t+1} + \frac{\gamma_4^t}{\mu_4}\|_2^2 \quad (S12)$$

$$x^{t+1} = \left(\mu_1 I + \mu_2 I + \mu_3 I + \mu_4 \Psi^T \Psi\right)^{-1}$$
$$\left[\mu_1 C^+\left(v^{t+1} - \frac{\gamma_1^t}{\mu_1}\right) + \mu_2\left(u^{t+1} - \frac{\gamma_2^t}{\mu_2}\right) + \mu_3\left(w^{t+1} - \frac{\gamma_3^t}{\mu_3}\right) + \mu_4 \Psi^T\left(z^{t+1} - \frac{\gamma_4^t}{\mu_4}\right)\right] \quad (S13)$$

$$C^+ = (C^T C)^{-1} C^T \quad (S14)$$

### B.2. Updating the duals
Dual variables are updated as:

$$\gamma_1^{t+1} = \gamma_1^t + \mu_1\left(Cx^{t+1} - v^{t+1}\right) \quad (S15)$$

$$\gamma_2^{t+1} = \gamma_2^t + \mu_2\left(x^{t+1} - u^{t+1}\right) \quad (S16)$$

$$\gamma_3^{t+1} = \gamma_3^t + \mu_3\left(x^{t+1} - w^{t+1}\right) \quad (S17)$$

$$\gamma_4^{t+1} = \gamma_4^t + \mu_4\left(\Psi x^{t+1} - z^{t+1}\right) \quad (S18)$$

### B.3. Data preprocessing and empirical parameter selection

For data preprocessing, all the measurements are normalized to 0 to 1 before feeding into the ADMM algorithm. The total energy of basis PSFs are normalized to be 1 and the coefficient maps are padded to match the measurement size and normalized to a maximum value of 1. The reconstruction $x^{t=0}$ and all the dual variables are initialized to zero.

Empirically, we find that the sum of basis PSFs has to be positive to enable the algorithm to converge; otherwise, the reconstruction may yield an image composed entirely of black pixels. For further clarification, consider the first update of the data fidelity term $v$. Given $x^{t=0} = 0$ and $\gamma_1^{t=0} = 0$, the update for $v^{t=1}$ follows:

$$v^{t=1} = \left(B^T B + \mu_1 I\right)^{-1} \left(B^T y\right) \tag{S19}$$

Since the denominator is always positive, the sign of $v^{t=1}$ depends on sign of the numerator $B^T y$, which is the summation of convolutions between the input measurement and the flipped basis PSFs. Given the input measurement, the fluorescence distribution is inherently positive. Therefore, the sign of $v^{t=1}$ depends on the summation of the basis PSFs. If this summation is negative, $v^{t=1}$ will be negative, leading $x^{t=1}$ to be negative. Consequently, the non-negativity constraint may dominate the optimization, hindering convergence. However, by ensuring a positive summation of the basis PSFs, we have empirically found that the non-negativity term can effectively enhance the reconstruction quality, as shown in Fig. S3.

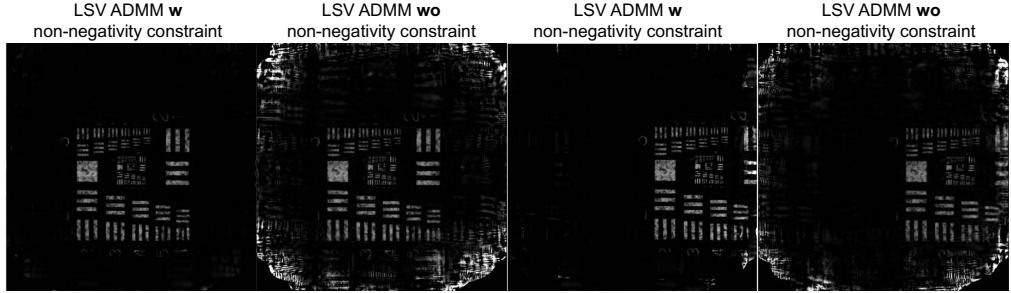

**Fig. S3. Comparison of LSV model based reconstruction with and without non-negativity constraint**. The reconstruction that incorporates the non-negativity constraint demonstrates improved image quality, particularly at the image edges.

## 3. ADDITIONAL DETAILS ON THE FORWARD MODEL

### A. Validation on forward model

To validate the accuracy of the forward model, we compare the experimental and simulated measurement using the same object, as shown in Fig. 2e in the main text. Initially, we imaged a phantom that composed of uniformly distributed 10 µm fluorescent particles with our computational multi-aperture miniscope. Concurrently, we acquired a table-top wide-field measurement to serve as the ground truth object. This ground truth object was subsequently resized to align with the miniscope's grid and input into the forward simulator to generate the corresponding simulated measurement, which highly matches to the measurement from our miniscope system as shown in Fig. S4. To evaluate the model accuracy, histograms of the intensity distributions from both measurements were plotted on standard and logarithmic scales. The histogram for the simulated data closely matches with the experimental data, demonstrating the forward model can accurately simulate realistic experimental measurements.

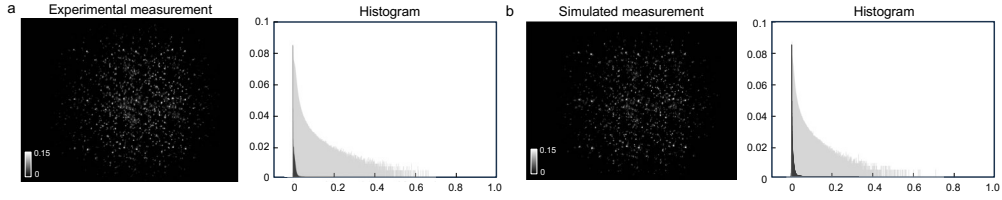

**Fig. S4. Comparison of experiment and simulated measurement.** The intensity distribution and its corresponding histograms of both experimental (a) and simulated measurements (b) are depicted: standard histograms are shown in black, with the y-axis representing the probability of each pixel intensity and the x-axis showing pixel intensity. Additionally, logarithmic histograms are plotted in gray to more clearly demonstrate variations in the intensity distribution at higher intensity levels.

### B. Noise model

The noise sources for a CMOS sensor typically include shot noise, readout noise, and fixed pattern noise. The intensity of a noisy pixel $g_i^n$ can be modeled as:

$$g_i^n = \alpha P(g_i^0/\alpha) + N(0, \beta) + \epsilon_i, \tag{S20}$$

where $g_i^0$ represents the intensity when pixel $i$ is clean. The shot noise is modeled by a Poisson distribution with parameter $\alpha$, readout noise is modeled as Gaussian distribution with variance $\beta$, and $\epsilon_i$ denotes the fixed pattern noise.

To calibrate this noise model, we first address the fixed pattern noise, which is a pixel-dependent constant caused by non-uniform reference voltage biases across the sensor pixels. This offset can be effectively estimated by averaging 100 frames of dark images. Subsequently, this offset is subtracted from all experimental measurements before reconstruction.

After removing this offset noise, the remaining noise components can be represented as mixed Poisson and Gaussian (MPG) noise:

$$g_i^{MPG} = g_i^n - \epsilon_i = \alpha P(g_i^0/\alpha) + N(0, \beta), \tag{S21}$$

To further calibrate these remaining noise components, we treat the Poisson noise as signal dependent Gaussian noise. This assumption is valid when the photon count exceeds five, making it a reasonable approximation for general cases. Following this approximation, the MPG noise can be modeled as pixel-dependent Gaussian noise, with the variance described by the following equation:

$$\sigma^2(g_i^{MPG}) = \alpha g_i^0 + \beta, \tag{S22}$$

where $\alpha g_i^0$ is the variance of the Poisson noise and $\beta$ is the variance of the Gaussian noise. This relationship implies that the noise level of measurement depends on $g_i^0$, which is linearly proportional to the object's fluorescence distribution. These noise parameters are then estimated with an established noise estimation algorithm [2] applied to a series of experimental measurements from various samples. Note that these parameters only related to the quantum efficiency and camera

digital gain, which are fixed during experiments. The resulting noisy measurement $g_n^i$ is given by:
$$g_i^n = g_i^0 + N(0, \alpha g_i^0 + \beta), \tag{S23}$$
where the calibrated noise parameter $\alpha$ ranging from $7.8109 * 10^{-5}$ to $9.6636 * 10^{-5}$ and $\beta$ varies between $1.3836 * 10^{-8}$ to $1.1204 * 10^{-7}$.

During the training of the network, the parameters $\alpha$ and $\beta$ are sampled from a uniform distribution within their calibrated ranges and the noise is sampled from Eq. (S22) and added to the measurements in an "online" fashion. This method ensures a comprehensive representation of the sensor's noise statistics, enriches the simulated data, and enables the synthetic measurements to closely approximate real imaging conditions.

### C. Synthetic data preparation

Our synthetic data contains 4,000 biological samples and 1,000 synthetic samples embedded with fluorescent particles.

The biological samples are sourced from several public repositories associated with published studies. Specifically, the dataset includes 1000 images derived from the 3D whole mouse brain vasculature dataset [3] and another 1000 images from the 3D whole mouse brain neuron dataset [4]. Initially, the raw 3D data are processed using the volume segmenter toolbox in Matlab to eliminate background and reduce noise. Subsequently, the cleaned volumes are randomly sliced at different angles to create 2D ground truth images. Given that the mouse brain size is comparable to our FOV, we maintain the physical size of the data by rescaling the images to match our effective pixel size. Additionally, we have collected 2,000 2D cell images of varying types, morphologies, and densities [5–7]. Unlike the 3D data, these cell images are generally smaller than our FOV and are thus not rescaled, preserving their original size.

Furthermore, we include 1,000 samples of synthetically generated fluorescent particles. The FOV for each synthetic sample follows a uniform random distribution $U[6.0\text{ mm}, 6.5\text{ mm}]$, and the emitter's diameter follows a uniform random distribution $U[8\ \mu\text{m}, 20\ \mu\text{m}]$. The emitter's intensity is set to the square of the diameter with a random scaling factor following a uniform distribution $U[0.8, 1.2]$. The emitter density per sample follows a uniform distribution $U[10, 50]$ (number of emitters / mm$^2$).

All biological and synthetic fluorescent samples constitute the ground truth and are subsequently sent into the forward model to synthesize the measurement. These measurements are subsequently normalized to a range of 0 to 1 and processed by the network for noise augmentation. To ensure consistency in the noise distribution across both simulations and experiments, we employ the same digital gain settings used during noise calibration and ensure the maximum pixel value for the experimental measurements saturates the camera.

## 4. ADDITIONAL DETAILS ON THE NETWORK IMPLEMENTATION

### A. Cropping and padding operation on the raw measurement

Due to the relative shift of the MLA to the sensor center, we first **crop** the the raw measurement from $3000 \times 4000$ pixels to $2886 \times 3500$ pixels pixels in order to align the measurement center with the PSF center and remove the highly aberrated information from the extreme edge of the measurements. Next, this processed measurement is zero-**pad**ded to $4200 \times 4200$ pixels for fast Fourier transform.

To verify the effectiveness of the use of **crop** and **pad** operations. We trained two networks: one incorporating all the mentioned crop and pad, and the other solely incorporating the crop operation for aligning the center. First, the comparison of reconstruction results for experimentally captured *C. elegans* from networks with and without crop and pad operations demonstrates that the crop and pad effectively reduces ghosting artifacts (Fig. S5(a)).

We further explain the ghosting artifact from the ERF perspective. ERF determines how the measurement areas contribute to the reconstruction at a given point. In the network trained without crop and pad, the presence of ghosting foci within the ERFs can lead to ghosting artifacts in the reconstruction, shown in Fig. S5(b). This issue arises when these ghosting foci erroneously gather information from measurement areas that correspond to sources in different locations. Consequently, even in the absence of a point source at the specified location in the reconstruction, these ghosting foci might still accumulate information from other regions, leading to the formation of a ghosting point. This explains why *C. elegans* from one part of the FOV may erroneously appear as a false duplicate in a different location in the reconstruction that is shown in Fig. S5(a).

Moreover, we perform a quantitative analysis of both networks using a simulated testing data set. The comparison of the MS-SSIM and PSNR for the reconstructions of this testing dataset is shown in Fig. S6. The median values are indicated by the red bars, demonstrating that the network trained with the crop and pad technique achieves better image quality and SNR.

### B. Implementation details on SV-FourierNet

This section details the network structure and number of trainable parameters of the SV-FourierNet.

Given that both measurement and object are real-valued signals in the object space, the measurements are initially transformed to frequency space through the *real* Fourier transform, which reduces computational and memory usage as compared to the full-complex Fourier transform. Subsequently, the measurement spectrum is multiplied with the multi-channel Fourier filters, enabling shift-variant blind deconvolution. To align with the reconstruction grid, the initial reconstructions are center cropped to $2400 \times 2400$ before being forwarded to the enhancement module.

The enhancement module is a modified Residual Channel Attention Network (RCAN)[8]. The initial reconstructions are down-sampled by the PixelUnshuffle module to reduce the spatial dimension while concurrently reallocating these information to different channels. The subsequent backbone consists of residual groups, each integrated with res blocks, skip connections and channel attention, where the skip connections facilitate high-resolution information flow, and the channel attention dynamically regulates the weights of channel-wise features. Finally, the outputs from the residual groups are subjected to up-sampling via the PixelShuffle module, ensuring the reconstruction's spatial dimensions are consistent with the object. The detailed structure is visualised in Table S2 and Fig. 3 in the main manuscript.

### C. Ablation study on the number of channels for SV-FourierNet

We choose 9 channels for the Fourier filters in SV-FourierNet based on the trade-off between performance and computational complexity.

First, to investigate the impact of the number of channels (rank) in the Fourier filters on the reconstruction quality, we trained SV-FourierNet with channel numbers of 1, 5, 9, 15 and 30 and plotted the corresponding training and validation loss curve, as shown in Fig. S7(a). As the number of channels increases, the loss converges to a lower value, indicating that additional Fourier filters can encapsulate more shift variant information, thus leading to more accurate reconstructions. Moreover, we test these networks with the simulated testing data and compare the MS-SSIM and PSNR, which are presented in Fig. S7(b). Both the MS-SSIM and PSNR increase with the number of channels, aligning with the trends observed in the loss curve. However, the improvements in both loss and metrics become smaller, especially when the number of channels exceeds 9, suggesting that 9 channels are sufficient to capture the shift-variant information in our

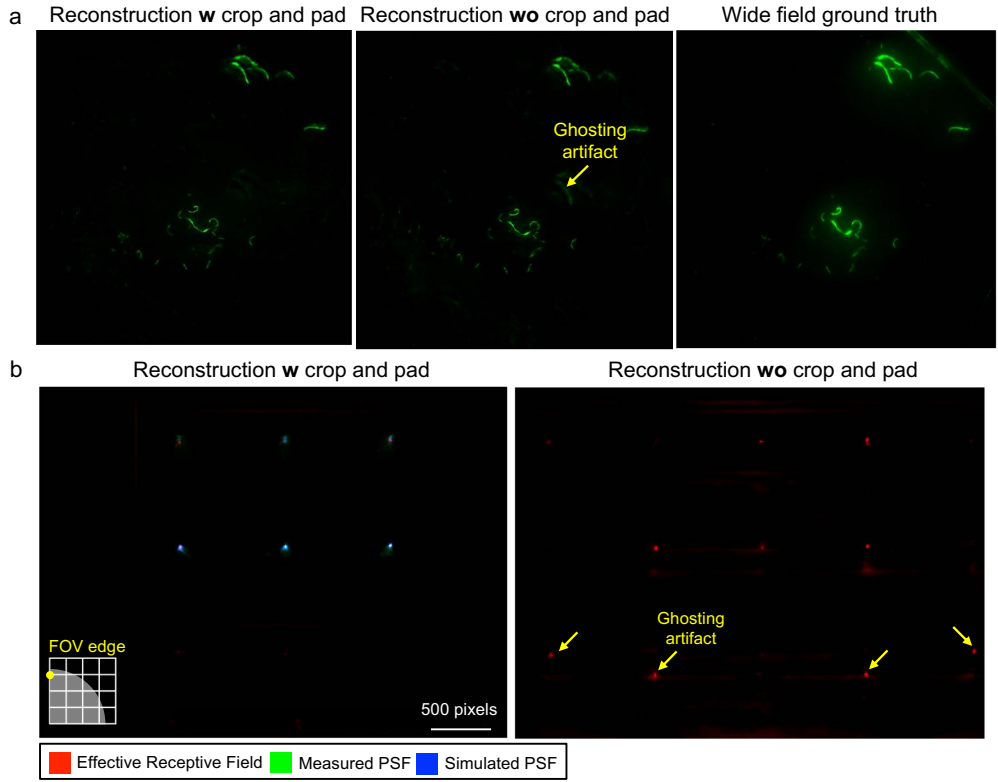

**Fig. S5. Comparison of reconstructions without (wo)/with (w) cropping and padding operations.** (a) Reconstructions of *C. elegans* from measurement with and without crop and pad, table-top wide field measurement serves as a reference. The network trained without crop and pad suffers from ghosting artifact. b: ERFs extracted from network wo/w crop and pad at a given position. The ghosting foci appeared in the ERF obtained from network without crop and pad corresponds to the ghosting artifact in the reconstruction.

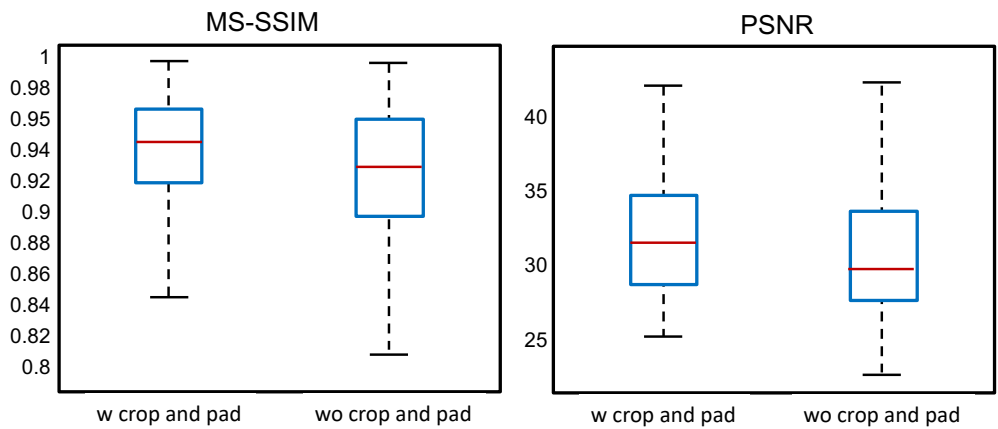

**Fig. S6. MS-SSIM and PSNR comparison for network trained w/wo crop and pad operations.**

system.

Fig. S8 provides a visual comparison of the reconstructions from both simulated and experimental measurements obtained using SV-FourierNet with varying numbers of channels. For a full FOV inspection, the image quality at the edge of the FOV significantly improves as the number of

10

**Table S2.** Parameter table for SV-FourierNet.

| Layer (type) | Output Shape | Param # |
|---|---|---:|
| rfft2 | [9, 4200, 4200] | 0 |
| Multi-channel Fourier layer | [9, 4200, 2100] | 79,380,000 |
| irfft2 | [9, 4200, 4200] | 0 |
| Center crop | [9, 2400, 2400] | 0 |
| PixelUnshuffle | [36, 1200, 1200] | 0 |
| Conv2d | [32, 1200, 1200] | 10,400 |
| Residual group 1 | [32, 1200, 1200] | 65,222 |
| Residual group 2 | [32, 1200, 1200] | 65,222 |
| Conv2d | [32, 1200, 1200] | 9,248 |
| Conv2d | [128, 1200, 1200] | 36,992 |
| PixelShuffle | [32, 2400, 2400] | 0 |
| Conv2d | [1, 2400, 2400] | 289 |
| Trainable params for Fourier blind deconvolution module | | 79,380,000 |
| Trainable params for Enhancement module | | 187,373 |
| Total trainable params | | 79,567,373 |

channels increases. In the zoomed-in regions, the spatial resolution accordingly enhances in both simulated and experimental reconstructions with an increasing number of channels. However, these improvements come at the cost of higher computational cost, thus leading to increased inference time.

Finally, we summarize the results and provide a quantitative comparison of different channel configurations in Table S3. Increasing the number of channels improves both PSNR and SSIM; however, this enhancement comes at the cost of larger model size, increased computational cost, and higher memory usage. After considering this trade-off, we opted for the network with 9 channels for the final implementation, striking a balance between performance and computational complexity.

**Table S3.** Performance metrics for different channel configurations

| # Channel | # Parameters | SSIM | PSNR | Speed (s) |
|---:|---:|---:|---:|---:|
| 1 | 9.0M | 0.8876 | 27.8677 | 0.0499 |
| 5 | 44.3M | 0.9295 | 30.8149 | 0.0578 |
| 9 | 79.5M | 0.9400 | 31.8595 | 0.0668 |
| 15 | 132.5M | 0.9422 | 32.0069 | 0.0813 |
| 30 | 264.8M | 0.9454 | 32.2874 | 0.1162 |

### D. Model selection for SV-FourierNet and MultiWienerNet

In this section, we present detailed loss and metrics curves for SV-FourierNet and MultiWienerNet over 48 hours of training. As depicted in Fig.S9(a), MultiWienerNet completes fewer epochs than SV-FourierNet during the same training time. This is because MultiWienerNet learns PSFs in the spatial domain and incurs additional computations to perform Fourier transforms to obtain

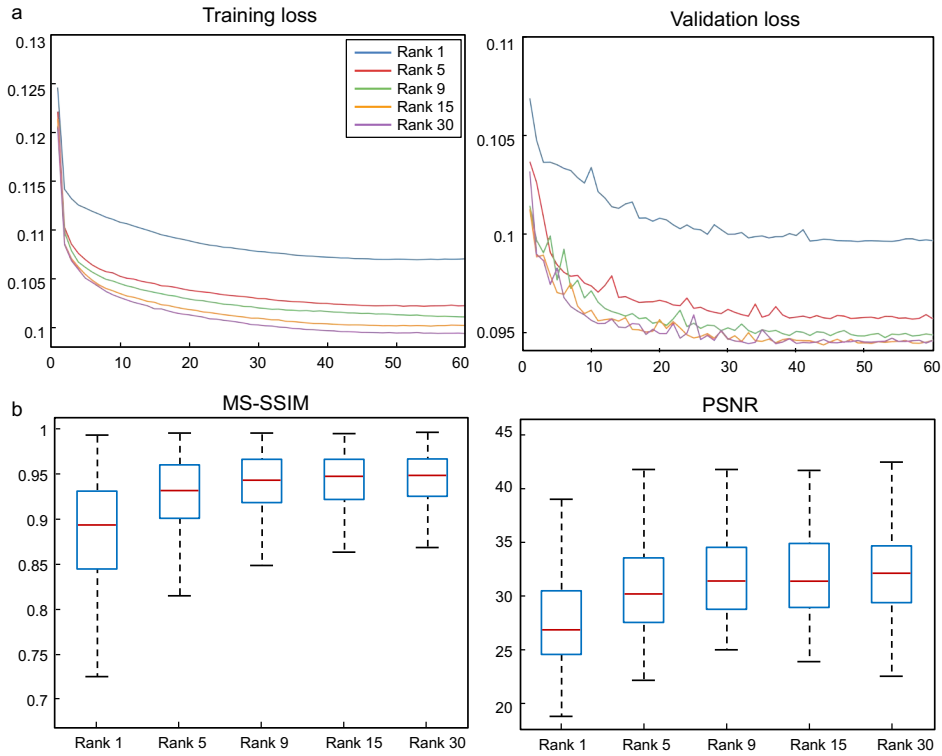

**Fig. S7. Loss curve and metrics evaluation for SV-FourierNet with different number of ranks.** (a) Training and validation loss curves. (b) The MS-SSIM and PSNR comparison.

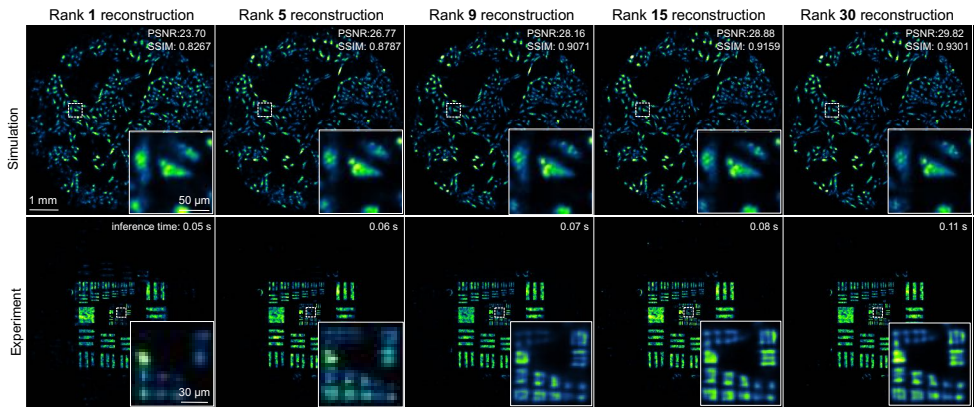

**Fig. S8. Reconstruction result comparison for SV-FourierNet with different number of channels.** The spatial resolution and image quality improve for both simulated and experimental reconstructions as the number of channels increases. However, this comes at the cost of increased inference time due to the higher computational load.

Wiener filters, resulting in fewer epochs within the allocated training time. However, both models' validation loss and MS-SSIM plateau before reaching the 48-hour, indicating convergence. The final models used in the benchmark studies are selected based on the best MS-SSIM scores, shown in Fig.S9(b). For SV-FourierNet, this occurs at 38 epochs and approximately 30 hours of training, while for MultiWienerNet, it is at 31 epochs and 38 hours.

12

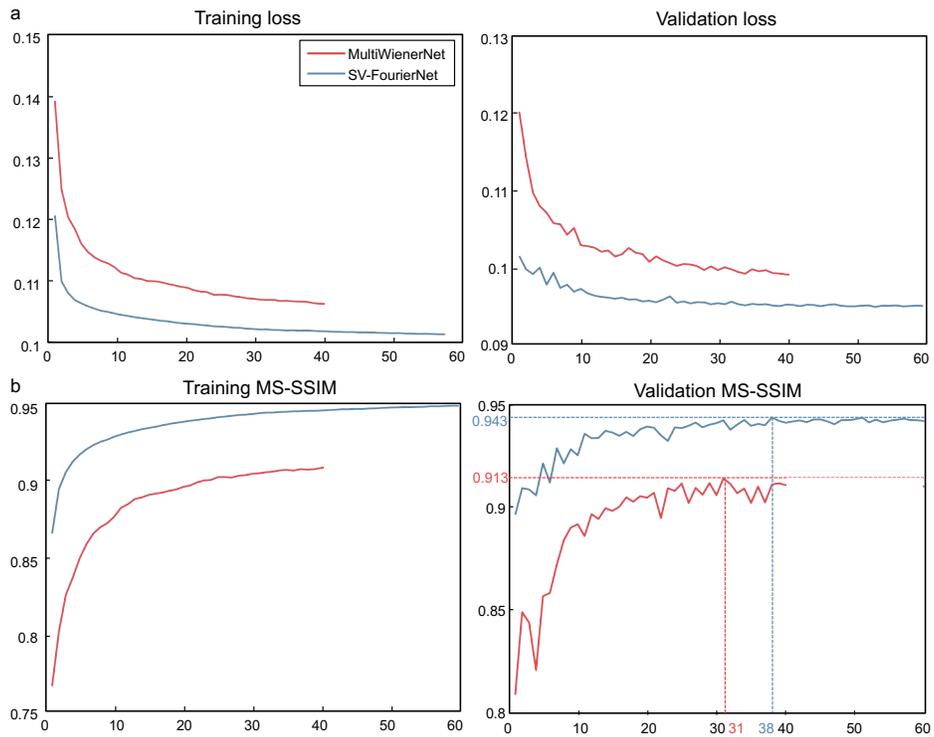

**Fig. S9. Comparison between SV-FourierNet and MultiWienerNet.** (a) The training and validation loss curve for SV-FourierNet and MultiWienerNet within 48 hours. (b) The training and validation MS-SSIM curve for SV-FourierNet and MultiWienerNet.

13

# 5. ADDITIONAL DETAILS ON IMPLEMENTATION AND EXPLANATION FOR EFFECTIVE RECEPTIVE FIELD

The effective receptive field (ERF) at a specific output pixel is defined as the region containing input pixels that significantly impact on that output unit. Studying the ERF offers a qualitative assessment of a network's implementation effectiveness. The concept of ERF has been extensively explored across various domains within computer vision, such as image classification[9], segmentation[10], and detection tasks[11]. In this work, we expand the application of ERF to deconvolution tasks and establish a correlation between ERF and physical optical systems.

## A. Detailed implementation of ERF

The ERF utilizes partial derivatives to indicate how input measurement impacts the final network reconstruction. Let $g$ represents the network input measurement, and $\widetilde{o}$ stands for the output reconstruction. The partial derivative $\frac{\partial \widetilde{o}_{x,y}}{\partial g_{u,v}}$ quantifies the effect of the input pixel at $(x,y)$ on the output pixel at $(u,v)$.

In general, the gradient back-propagation during the training process is guided by the chain rule, which can be mathematically expressed as:

$$\frac{\partial l}{\partial g_{u,v}} = \sum_{x',y'} \frac{\partial l}{\partial \widetilde{o}_{x',y'}} \frac{\partial \widetilde{o}_{x',y'}}{\partial g_{u,v}} \tag{S24}$$

where $l$ stands for the loss function. $(x', y')$ are indices corresponding to units in intermediate layer while back-propagation. To calculate the partial derivative for a specified point $\widetilde{o}_{x_0,y_o}$ relative to the input, the loss gradient $\frac{\partial l}{\partial \widetilde{o}_{x_0,y_0}}$ is set to 1, while all others $\frac{\partial l}{\partial \widetilde{o}_{x,y}}$ are set to 0. This gradient back-propogation is facilitated by the autograd function in PyTorch, allowing for generating a heatmap map that illustrates how input measurement influences any specific point $(x_0, y_0)$ in the reconstruction.

Note that the ERF depends not only on the learned weights of the network but also on the input. To render the ERF input-independent, we averaged the results across the entire testing dataset.

## B. Quantitative ERF analysis

To quantitatively assess the network's deconvolution performance with ERF, we developed metrics to quantitative evaluate and compare the ERFs of SV-FourierNet and MultiWienerNet.

First, as illustrated in the manuscript, the positive ERF aligns with the physical PSFs. By plotting the line profiles of the foci in the positive ERF, we compare the spatial resolution of both networks, as shown in Fig. S10. SV-FourierNet consistently exhibits a smaller FWHM compared to MultiWienerNet, indicating higher spatial resolution due to less influence from nearby measurement signals on the reconstruction point. This is further supported by the resolution target reconstructions in Fig. S24.

Secondly, we calculated the Signal to Background Energy Density Ratio (SBER) on the ERFs to assess the network's background energy leakage. This ratio is calculated as follows:

$$\text{SBER} = \frac{\frac{1}{N} \sum_n \text{ERF}_n}{\frac{1}{M} \sum_m \text{ERF}_m}, \quad \text{where } N \in \text{Signal Region and } M \in \text{Background Region} \tag{S25}$$

Here, N and M represent the number of pixels in the signal and background regions, respectively. The signal regions are defined as dashed squares (60 x 60 pixels) containing the 3 x 3 foci, with the remaining area constituting the background.

At the center FOV, SV-FourierNet achieves a SBER of 150.14, significantly surpassing MultiWienerNet's 42.16. At the edge FOV, SV-FourierNet also maintains a higher SBER of 72.32, compared to 32.57 for MultiWienerNet. These consistently higher SBER for SV-FourierNet indicate a higher energy concentration at the foci relative to the background, which is visually corroborated by the log intensity comparison of ERF shown in Fig. S10(b). Conversely, the lower SBER observed in MultiWienerNet suggests energy leakage into the background, potentially resulting in a more noisy reconstructions with ghosting artifacts.

## C. ERF evolution over time

To illustrate the development of the ERF throughout the training process, we plotted the SV-FourierNet's ERFs at various epochs. As depicted in Fig. S11, the ERFs progressively align with

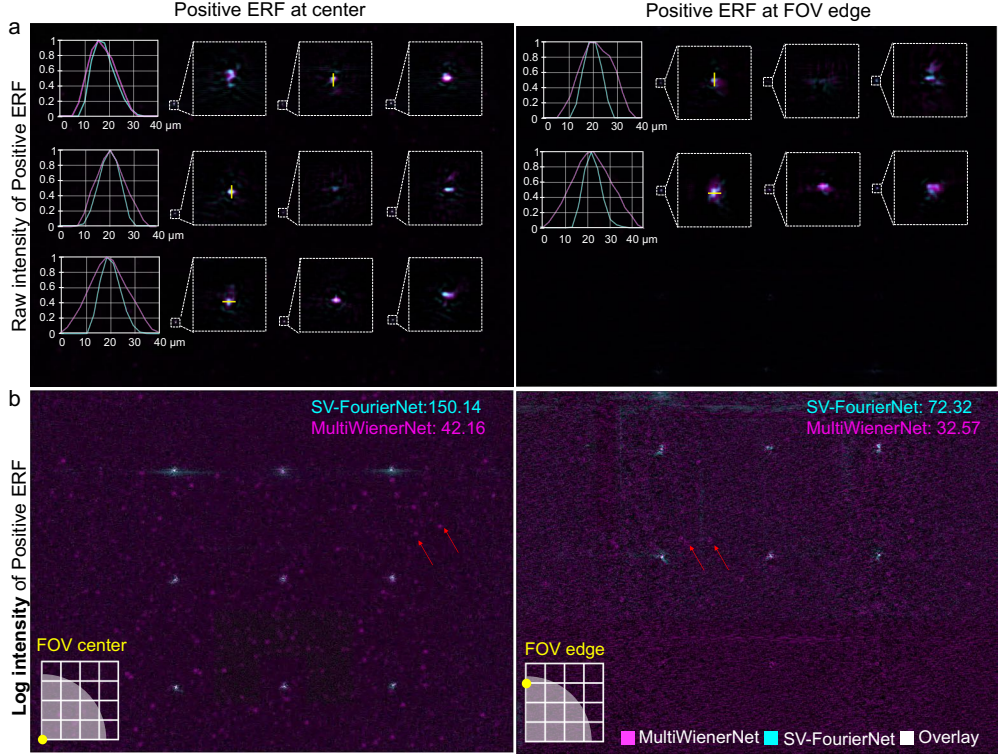

**Fig. S10. Quantitative positive ERF comparison for SV-FourierNet and MultiWienerNet.** (a) Overlay of the ERF obtained with SV-FourierNet and MultiWienerNet at both the center and edge of the FOV. Line profiles extracted from the ERFs reveal that SV-FourierNet maintains a tighter focus compared to MultiWienerNet, indicating superior spatial resolution. (b) Log intensity analysis of the ERFs shows a higher background level in MultiWienerNet compared to SV-FourierNet. SV-FourierNet demonstrates greater energy concentration at the foci, whereas MultiWienerNet exhibits energy leakage into the background, potentially leading to ghosting artifacts in reconstructions.

the physical PSFs over time, demonstrating the network's ability to learn the inverse mapping from extensive training data. For a clearer visualization, line profiles from three foci extracted from both center and edge FOV ERFs, demonstrate increasingly focused points across epochs. This enhancement corresponds to the network's gradual improvement in reconstruction accuracy and resolution. The steady convergence of the ERFs to the physical PSFs serves as a crucial indicator of the network's effective learning and successful implementation.

### D. Interpretation for negative ERF

The negative ERF is defined by the negative part of the partial derivative of a specific point in the reconstruction relative to the measurement. It indicates the regions within the measurement that inversely affect the point in the reconstruction, highlighting the network's capability to suppress aberrations and demultiplex signals from sources with ambiguous PSFs.

Initially, we compare the positive and negative ERFs at different FOV positions, as shown in Fig. S12(a). The negative component consistently surrounds the tight foci of the positive component, demonstrating the network's capability to suppress the signal around the sharp foci, thereby improving the resolution.

Moreover, compared to the positive component, the negative component at the FOV edge displays additional foci that are displaced from the PSF by a distance equal to one complete cycle of the repeating pattern caused by MLA, as indicated by the red dashed box. We hypothesize that these additional foci in the negative ERF demonstrates the network's capability to demultiplex the information arising from points in the FOV that suffer from PSF ambiguity. Consider a scenario where our FOV includes two point sources: one at the center (cyan) and another at the edge

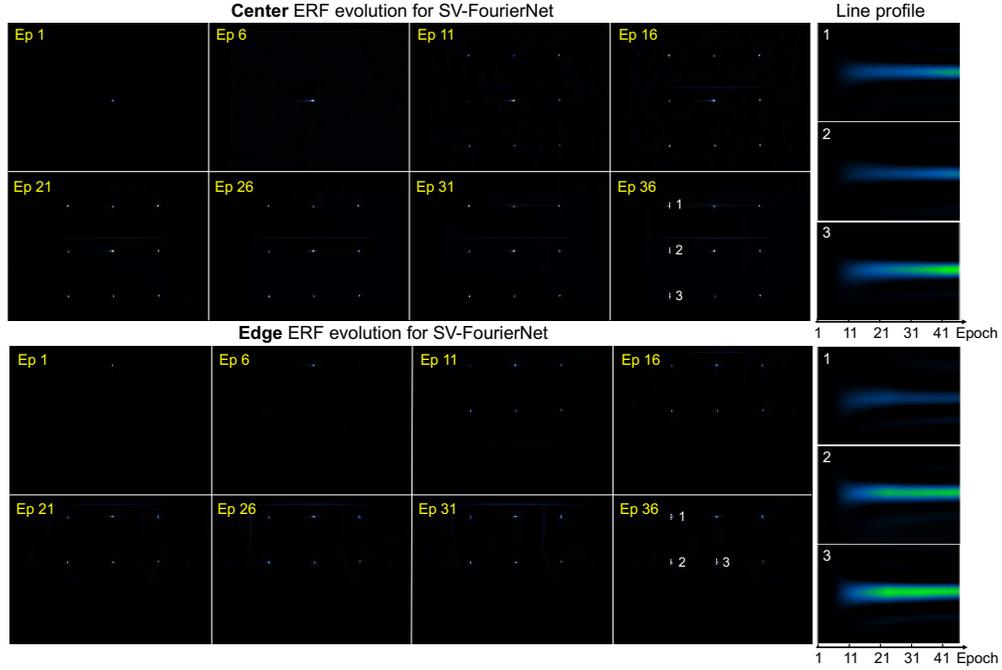

**Fig. S11. Positive ERF evolution during training process.** The ERF at the center and edge learned during the training process shows progressive convergence to the physical PSFs. Line profiles from three foci illustrate the evolving localization accuracy throughout the training period.

(magenta), as shown in Fig. S12(b). The PSF from the edge point overlaps with that from the center (the overlapped 2 x 3 PSF foci is displayed in white). The negative ERF obtained from the edge point corresponds to the distinct, non-overlapping foci in the measurement sourced from the center point (labeled with cyan), showcasing the network can distinguish and demultiplex the center point's signal from the edge point.

This analysis of the negative ERF demonstrates the network's effectiveness in suppressing aberrations and resolving points with ambiguous PSFs, effectively enhancing the quality of the reconstruction and facilitating demultiplexing.

Further emphasizing this capability, Fig. S13(a,b) presents simulation results that highlight the challenge of distinguishing between highly similar measurements from periodic signals. Despite this, SV-FourierNet accurately distinguishes these cases and recovers the objects correctly. Moreover, in situations where measurements for two different digits, separated by the PSF periodicity, strongly overlap, the network still manages to successfully recover both digits. This underscores SV-FourierNet's ability to resolve and demultiplex overlapping signals.

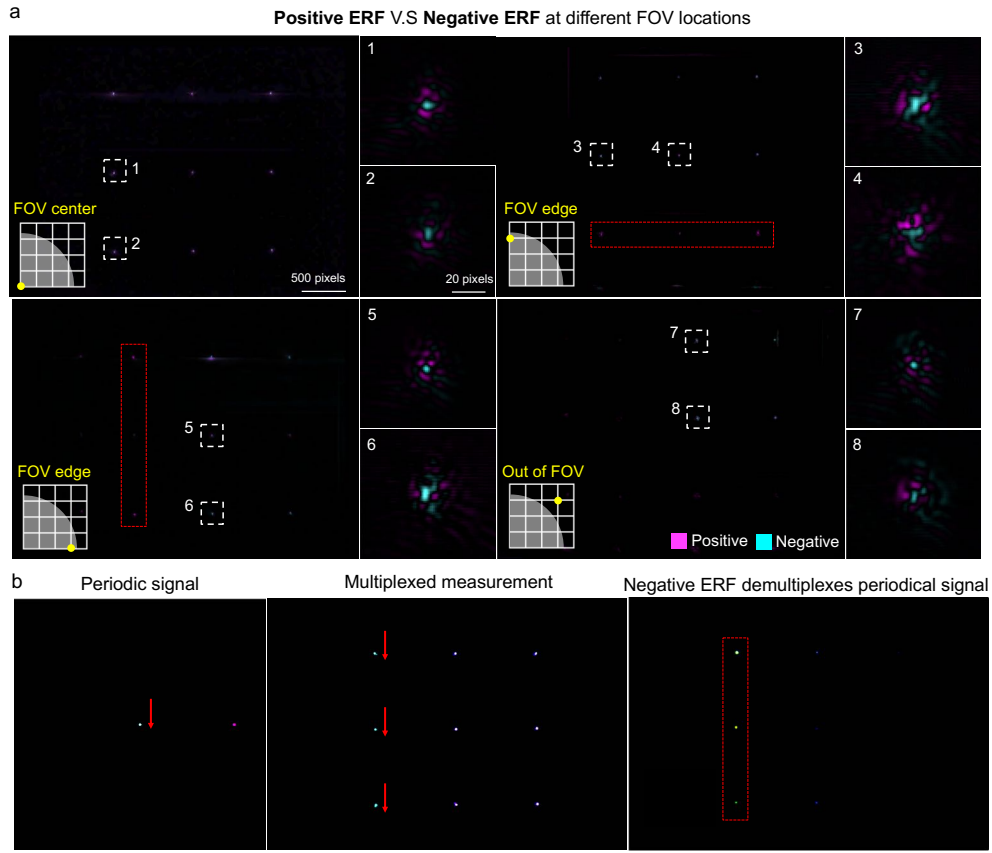

**Fig. S12. Negative ERF.** (a) Comparison of the positive and negative part of ERF at different FOV positions. The negative ERF surrounds the positive ERF, demonstrating the network's capability to suppress aberrations and improve the spatial resolution. (b) Periodic point sources with ambiguous PSFs results in multiplexed measurement, where magenta and cyan represent signals from the edge and center points, respectively. The negative ERF derived from the edge point isolates the distinct signal from the center point, illustrating the network's capability to effectively distinguish and demultiplex the center point's signal from that at the edge.

17

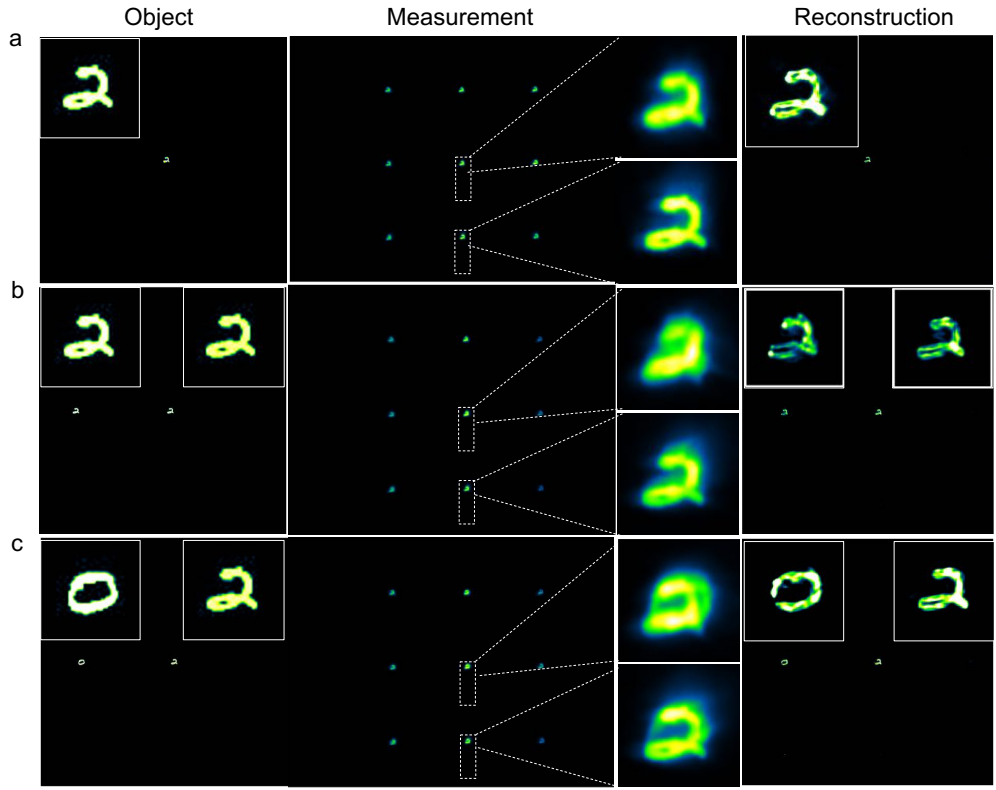

**Fig. S13. Demultiplexing overlapped objects with SV-FourierNet.** (a) Simulated measurement of digit 2 placed at the FOV center. Compared to the measurement, the SV-FourierNet accurately reconstructs the digit with enhanced contrast and resolution. (b) Two identical digits, separated by a distance that matches the PSF's periodicity, exhibit highly overlapped signals in the simulated measurement. Nevertheless, the SV-FourierNet can distinguish the components from both locations and reconstructs both digits. (c) Two different digits, also separated by a distance equivalent to the PSF's periodicity. The simulated measurement shows the digit 0 are highly overwhelmed by the signal from digit 2. Despite this, the SV-FourierNet can disentangle and accurately recover both digits.

## 6. ADDITIONAL DETAILS ON THE LEARNED FOURIER FILTERS

### A. Visualization of Fourier filters

The SV-FourierNet utilizes multiple learnable Fourier filters to adeptly manage the complex, non-local, and spatially-varying array PSFs. To illustrate the Fourier filters, we extract complex-valued Fourier weights from the multi-channel Fourier layer comprising nine channels, each with dimensions 4200 × 2100 pixels. For spatial domain visualization, we perform a real-valued inverse Fourier transform on these weights, resulting in filters of 4200 × 4200 pixels. These filters are then center-cropped to 3000 × 4000 pixels to match the measurement size. Fig. S14 showcases these Fourier filters, each extending across the entire sensor area, highlighting the network's global receptive field. A closer inspection through the zoomed-in sections of the central foci from different filters reveals distinct characteristics unique to each, highlighting the filters' shift-variant characteristics.

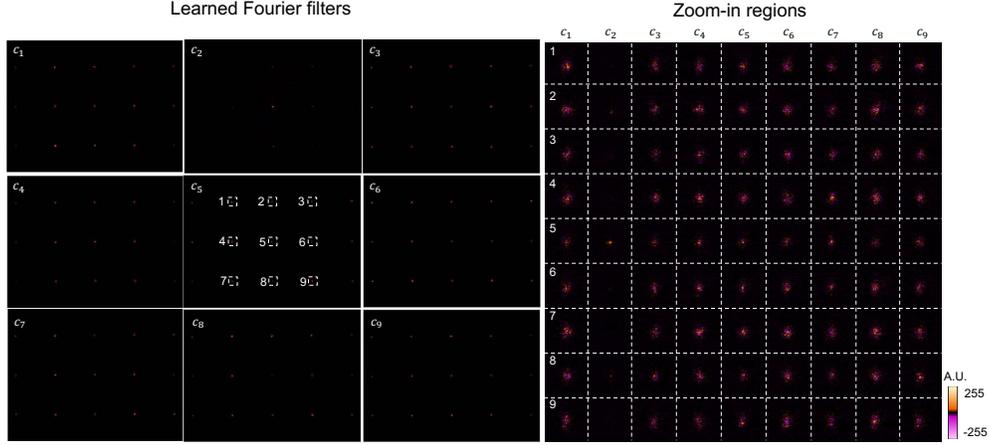

**Fig. S14. Visualization of effective spatial domain filters.** Inverse Fourier transform of the learned Fourier filters across nine different channels, as indexed by $c_i$, where $i = 1, 2, \cdots, 9$. The zoomed-in regions from the nine dashed box showing distinct foci features across different channels.

### B. Interpretation of the learned Fourier filters

We hypothesize that the learned Fourier filters encapsulate a low-dimensional projection of the range space associated with the shift-variant PSF, analogous to TSVD, which provides a low-rank approximation of the shift variant physical model that minimizes the least squares error. To validate this hypothesize, we compare our learned Fourier filters with the pseudo-inverse of the basis PSFs derived via TSVD from experimentally calibrated PSFs. Unlike the spatially-varying imaging model described in our manuscript, which decomposes PSFs separately for each ML, we globally approximate the shift variant PSFs by registering all the calibrated PSFs and performing TSVD directly on these registered raw measurements. The TSVD can be mathematically expressed as:

$$H(p,q;x,y) \approx \sum_{k=1}^{9} M_k(x,y) B_k(p,q), \tag{S26}$$

where $H(p,q;x,y)$ represents the registered calibrated PSFs, and $B_k(p,q)$ denotes the basis PSFs, which projects the physical shift variant PSFs into a lower dimensional orthonormal basis. $M_k(x,y)$ is the coefficient map, describing the relative weights of the basis PSFs to construct the physical PSF at point $(x,y)$.

Subsequently, we extract the basis PSFs $B_k$ and compute their pseudo-inverse with the following equation:

$$B_k^H = \text{ifft}\left(\frac{\text{fft}(B_k)^*}{|\text{fft}(B_k)|^2 + \lambda}\right), \tag{S27}$$

where $\lambda$ is a regularization term to suppress the noise amplification when $|fft(B_k)|$ is small. The results are shown in Fig. S15. These filters demonstrate a high degree of similarity to the learned

19

Fourier filters, each displaying a 3 × 5 array of foci. The detailed examination of the zoomed-in views from the nine dashed boxes reveals unique aberration patterns, akin to those observed in the learned foci, with each characterized by tight foci with distinctive aberrated patterns.

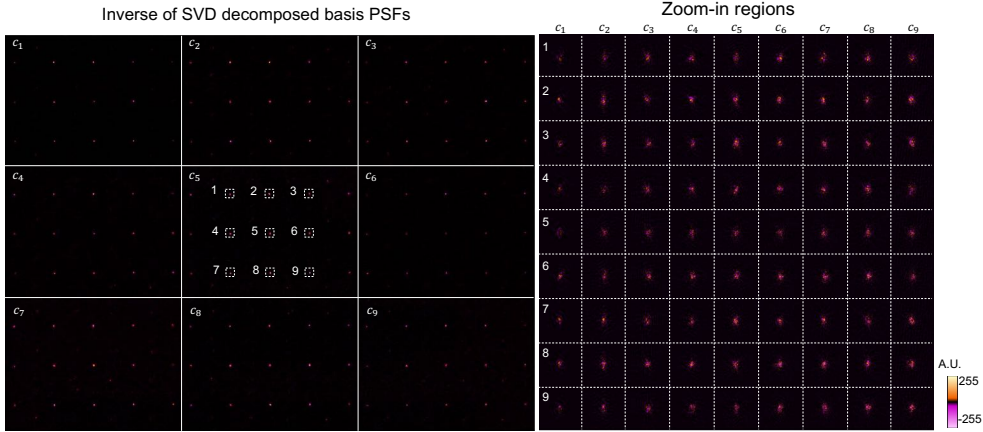

**Fig. S15.** Pseudo-inverse filters derived via SVD from calibrated shift-variant physical PSFs. These inverse filters show a high similarity to the learned Fourier filters, each displaying a 3 × 5 array of foci. Zoomed-in views from the nine dashed boxes highlights unique aberration patterns.

**C. Comparison of the learned Fourier filters with the learned Wiener filters**

SV-FourierNet, which learns a global decomposition of the inverse mappings of the physical PSFs, exhibits higher fidelity than the locally biased physical PSFs learned from MultiWienerNet.

Initially, to validate the effectiveness of the trained MultiWienerNet, Fig. S16(a) compares the initialized physical PSFs, experimentally captured from different viewpoints, with the learned PSFs. The learned PSFs exhibit increased noise levels, agreeing with the results shown in [12]. Subsequently, we compute the learned Wiener inverse filters based on Eq. (S27), where $B_k$ is replaced with the learned PSFs and $\lambda$ is replaced with the learned regularization constant. Fig. S16(b) shows these learned Wiener filters. For full measurement inspection, the learned Wiener filters suffer from a markedly higher noise level compared to the learned Fourier filters. In the zoom in regions, the learned Wiener filters are biased by aberrations and truncation patterns from the initialized local physical PSFs.

Moreover, we compare the initial reconstructions of an experimentally captured resolution target placed at the edge of the FOV from SV-FourierNet and MultiWienerNet, before and after learning. The initial reconstruction is calculated as:

$$\widetilde{o_0} = \text{crop}\left(\text{ifft}\left(W \odot \text{fft}(g)\right)\right), \tag{S28}$$

where $W$ is equal to the Fourier weights directly extracted from SV-FourierNet or $W$ is equal to the Fourier weights of the learned Wiener filters. We crop only the central 2400 × 2400 pixels to match the size of the object.

Fig. S17 shows the initial reconstructions before learning. Compared to the initial reconstruction from the Fourier filter initialized with a uniform distribution, Wiener filters initialized with physical PSFs exhibit similar resolution. This is because the initialized physical PSFs are locally biased to specific positions where they were captured. These filters are most effective at deconvolving measurements at those positions but are much less effective when the object is placed elsewhere due to the shift-variant nature of the system.

Fig. S18 shows the initial reconstructions after learning. Both reconstructions exhibit improved contrast and spatial resolution compared to those before learning. The initial reconstruction from learned global Fourier filters from SV-FourierNet demonstrates higher resolution and less noise compared to MultiWienerNet. The resolution in the initial reconstruction corresponds with the final reconstruction shown in Fig. S24.

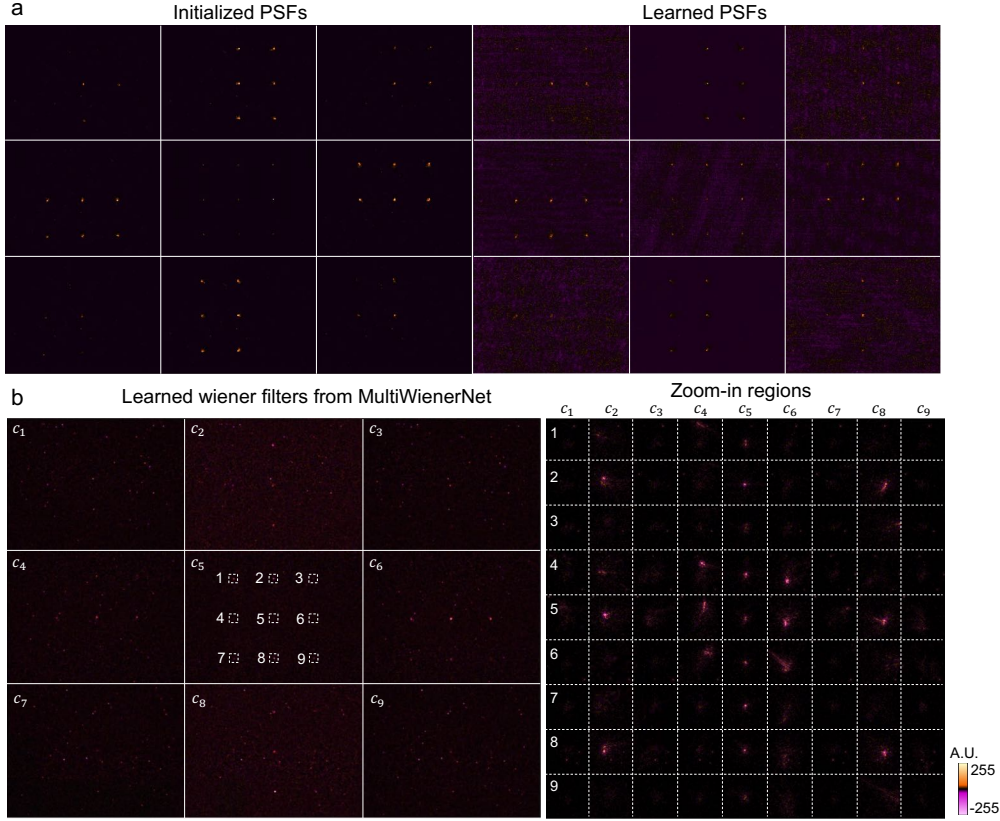

**Fig. S16.** (a) Comparison between the initialized physical PSFs and the learned PSFs from the MultiWienerNet. (b) The learned Wiener filters derived from the learned PSFs and regularization term, exhibiting a higher noisy level and are biased by the initialized local physical PSFs.

### D. Ablation study on physical PSF initialization and learning in Fourier/Spatial domain

To further investigate the impact of initialization and learning domain on network performance, we compare trained networks under identical configurations, varying only the presence or absence of physical PSF initialization and the learning domain, either spatial or Fourier. MultiWienerNet employs spatial domain learning with physical PSF initialization, denoted as "Spatial with PSF init" in subsequent discussions. Conversely, SV-FourierNet utilizes Fourier domain learning without PSF initialization, referred to as "Fourier without PSF init". To complete the ablation study, we introduce two additional networks: SV-FourierNet initialized with Wiener filters (Fourier with PSF init) and MultiWienerNet initialized with a uniform distribution (Spatial without PSF init).

Initially, we compare the loss and MS-SSIM curves of these networks trained for an equal number of epochs (Fig. S19). Regardless of whether the physical layer is initialized with a physical PSF or a uniform distribution, frequency-based learning consistently converges to lower loss and higher MS-SSIM compared to spatial learning. This highlights the advantages of Fourier domain learning over spatial domain. Furthermore, within the same learning domain, initializing the physical layer with physical PSFs consistently yields inferior results, indicating that Wiener filters initialized with physical PSFs provide a less favorable starting point for network training. Quantitative analysis on the testing datasets corroborates these trends, as summarized in Table S4.

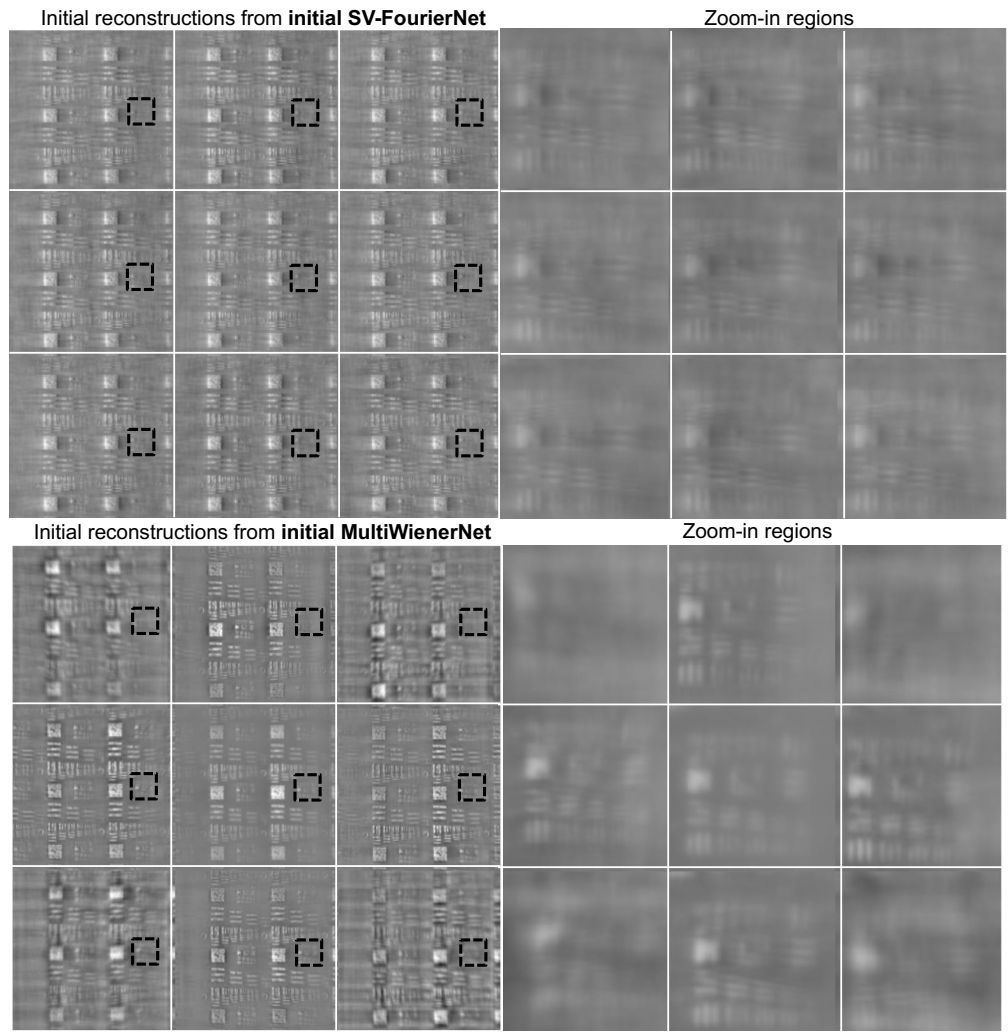

**Fig. S17.** Initial reconstructions obtained from SV-FourierNet, initialized with a uniform distribution, and MultiWienerNet, initialized with physical PSFs calibrated at different FOV positions. Zoomed-in regions reveal that these initial reconstructions exhibit similar resolution.

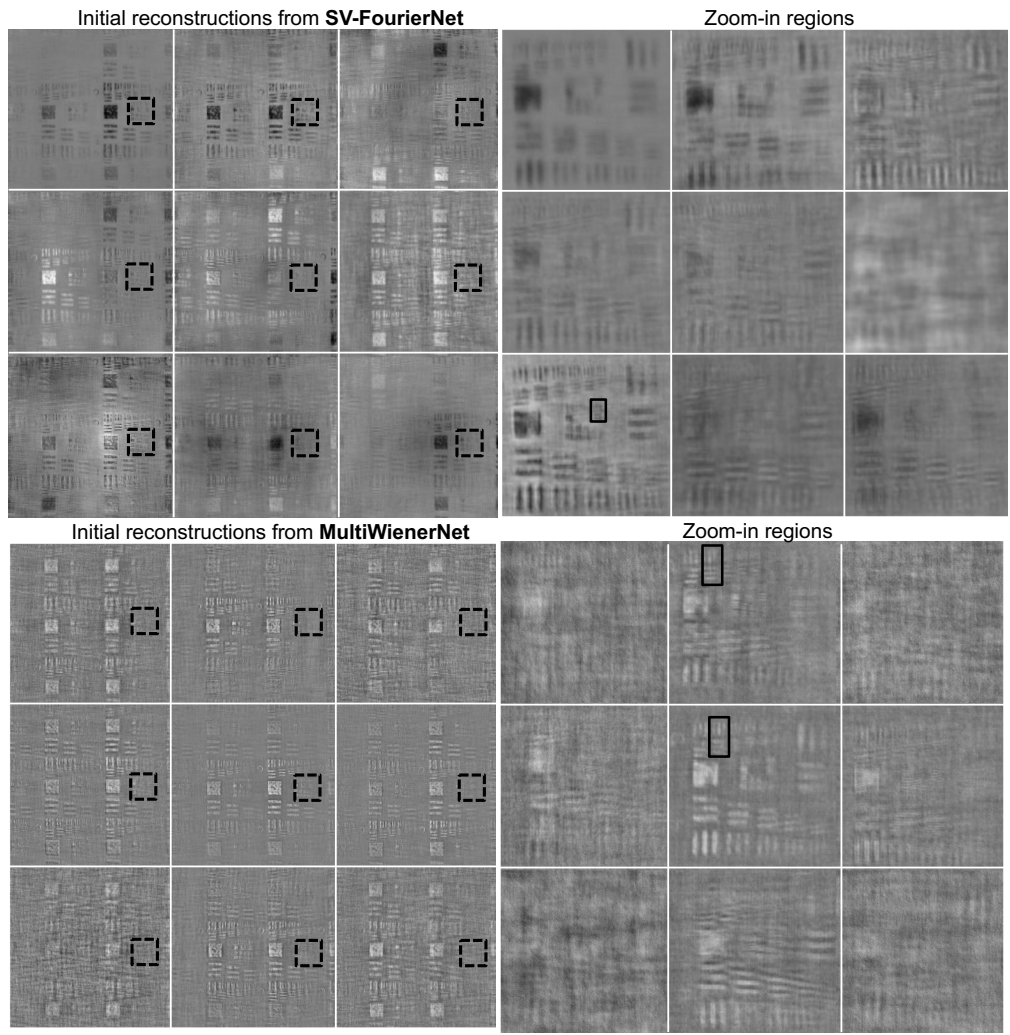

**Fig. S18.** Initial reconstructions from the learned SV-FourierNet and MultiWienerNet. The initial reconstructions from the SV-FoufierNet exhibit a higher spatial resolution and less noise level.

23

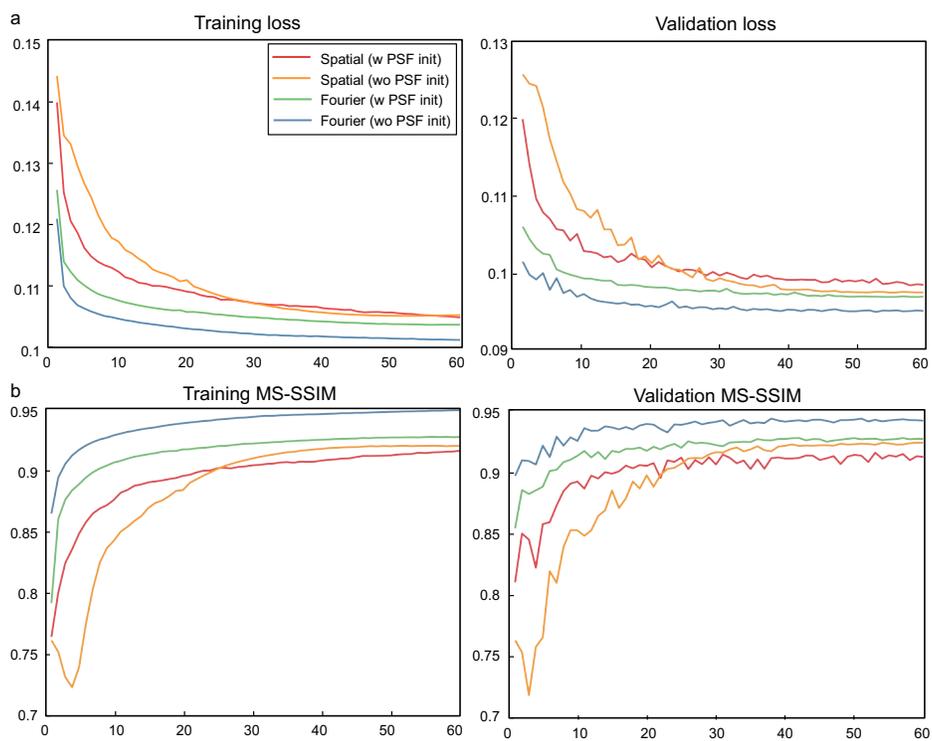

**Fig. S19.** Comparison of loss and MS-SSIM curves for networks with varied initialization conditions and learning domains during training.

**Table S4.** Quantitative evaluation w/wo physical PSF initialization and learning in Fourier/spatial domain

|  | **Fourier Domain Learning** | **Spatial Domain Learning** |
|---|---|---|
| **Initialized w PSFs** | SSIM: 0.9138 ⎮ PSNR: 29.53 | SSIM: 0.8995 ⎮ PSNR: 29.10 |
| **Initialization wo PSFs** | **SSIM: 0.9400** ⎮ **PSNR: 31.86** | SSIM: 0.9136 ⎮ PSNR: 29.10 |

To further illustrate the results, we visualize learned inverse filters and initial reconstructions of these new trained networks. Initial comparisons between results from SV-FourierNet ("Fourier without PSF initialization") [Fig. S14(a) and S18] and "Spatial without PSF initialization" (Fig. S20(a)), reveal that inverse filters learned blindly in the spatial domain can also capture a global decomposition of shift-variant filters but with less focused foci. This results in significant resolution degradation in initial reconstructions, highlighting the advantages of frequency domain learning over spatial domain learning. The inferior performance of spatial domain learning is attributable to the additional computational steps it requires. These extra steps not only increase the computational load and decrease network efficiency but also intensify approximation errors and noise accumulation, which can blur or distort sharp and sparse foci, compromising image quality.

To further evaluate the impact of initialization, we compare results from "Fourier without PSF initialization" [Fig. S14(a) and S18] to "Fourier with PSF initialization" (Fig. S20(b)). The inverse filters and initial reconstructions from the latter exhibit severe noise, attributed to our system's highly ill-conditioned nature, marked by severe sensor truncation and shift-variant degradation, which render Wiener deconvolution ineffective. This comparison underscores that prior information, such as the physical PSFs at predefined spatial locations, does not invariably enhance outcomes; rather, it can introduce biases related to local aberrations and inherent noise, trapping the network in local minima. Conversely, the network that learn directly from data can potentially learn more effective global Fourier inverse filters, yielding markedly better spatial resolution and overall image quality than those with physical PSFs initialization.

Finally, Fig. S20(c) presents a comparison of the final reconstructions of a resolution target at the FOV edge for networks under the four configurations previously discussed. Frequency domain learning consistently outperforms spatial domain learning in reconstruction quality under identical initialization conditions. Furthermore, compared to networks that learn directly from data, those with PSF initialization exhibit increased noise and hallucinated backgrounds in their reconstructions.

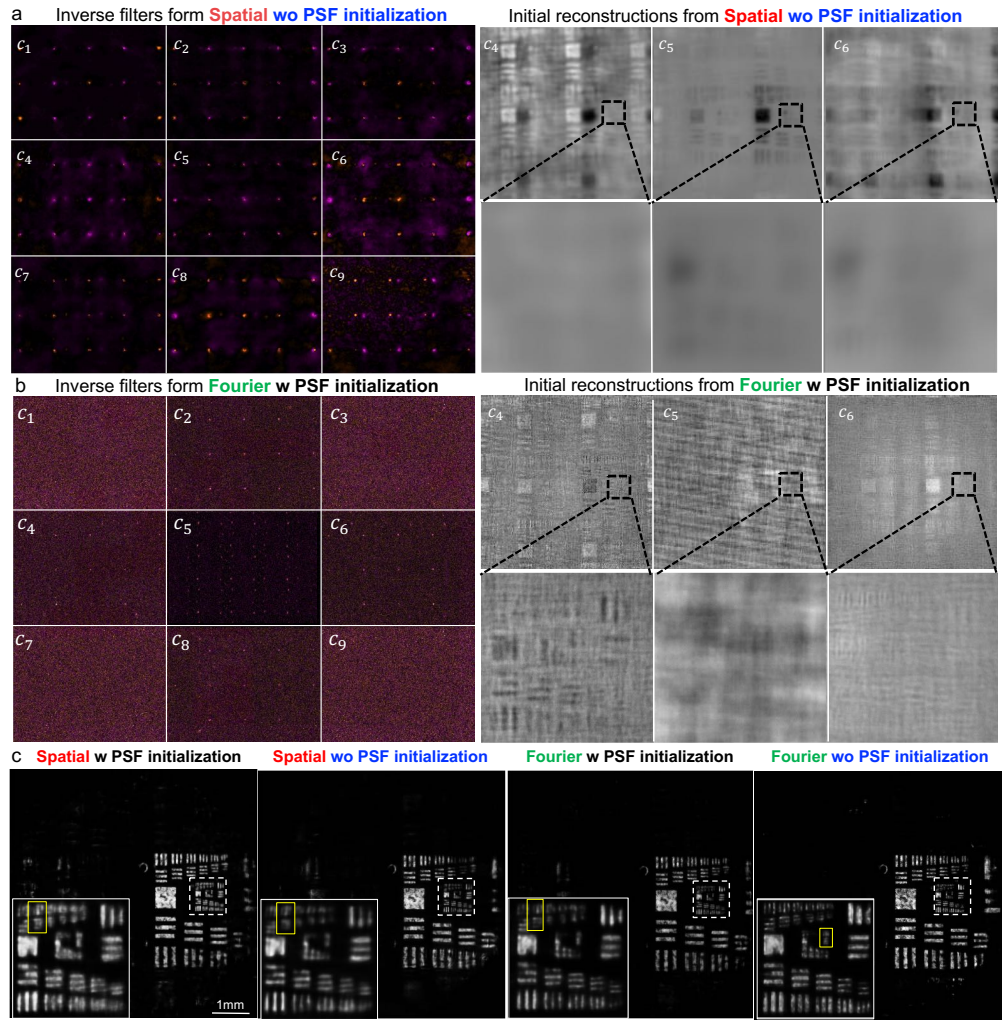

**Fig. S20.** (a,b) Inverse filters, along with initial reconstructions from channels $c_4$, $c_5$, and $c_6$, of networks trained in spatial domain without PSF initialization and trained in Fourier domain w PSF initialization. (c) Final reconstruction of the resolution target with different network configurations.

## 7. ADDITIONAL DETAILS ON BENCHMARK NETWORKS

The detailed network structures for the benchmark networks, including MultiWienerNet [12] and CM$^2$Net [13], are shown in Figure S21(a,b). To ensure a fair comparison, MultiWienerNet utilizes the same enhancement network as that of SV-FourierNet. The measurement is initially deconvolved with multiple Wiener filters, which are initialized with PSFs from different viewpoints for shift-variant deconvolutions. The resulting intermediate reconstructions are then processed through the RCAN, which synthesizes information across channels and enhances the reconstruction results. The gradients are back-propagated to simultaneously update both the parameters in the RCAN and the shift-varying PSFs.

CM$^2$Net is composed of view demixing, view synthesis, and enhancement modules. These modules share a uniform network architecture, which includes 20 standard Residual blocks (Res-blocks) followed by a convolution layer and a Sigmoid activation function.

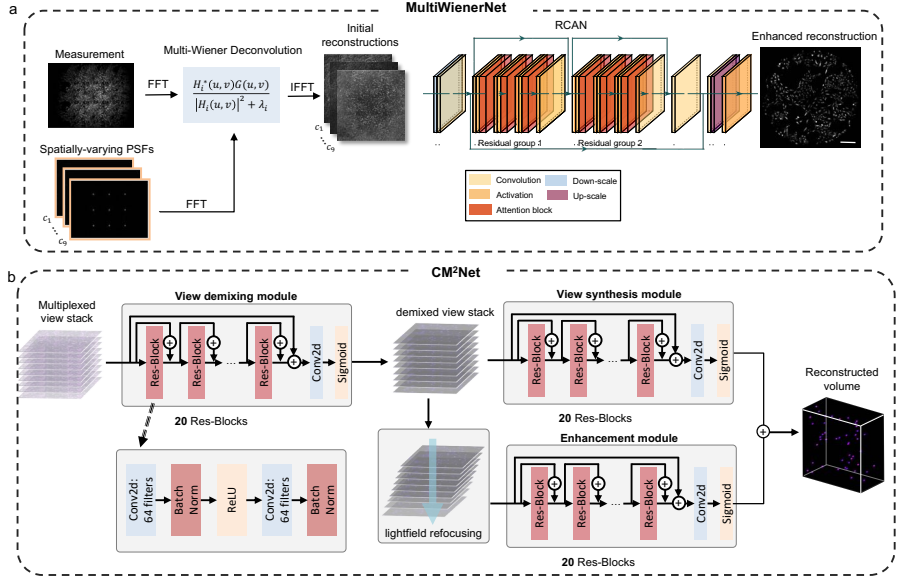

**Fig. S21.** Detailed network structure for MultiWienerNet and CM$^2$Net.

In addition to the reconstruction results presented in the manuscript, we show additional simulation and experiment results in Fig. S23. Moreover, we conducted a quantitative analysis for the benchmark comparison. All networks were tested on the same simulated dataset, and their performance was evaluated using Multi-Scale Structural Similarity Index (MS-SSIM) and Peak Signal-to-Noise Ratio (PSNR) as metrics. Our SV-FourierNet demonstrated superior image quality, achieving the highest PSNR among the compared networks, as shown in Fig. S22.

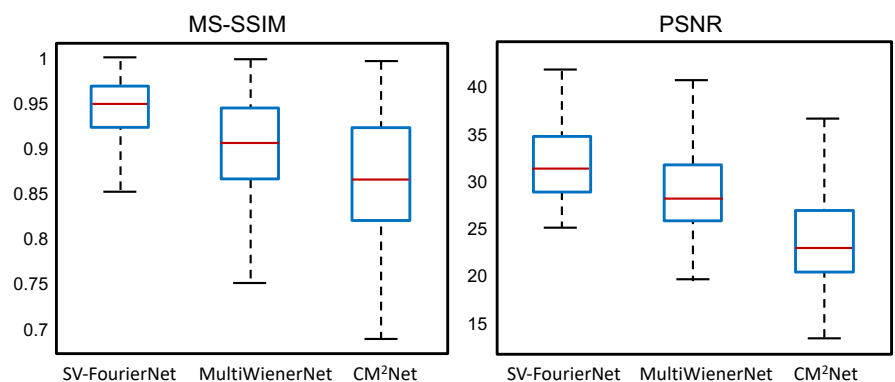

**Fig. S22. Quantitative analysis of SV-FourierNet, MultiWienerNet and CM$^2$Net.** The SV-FourierNet achieves best image quality and PSNR.

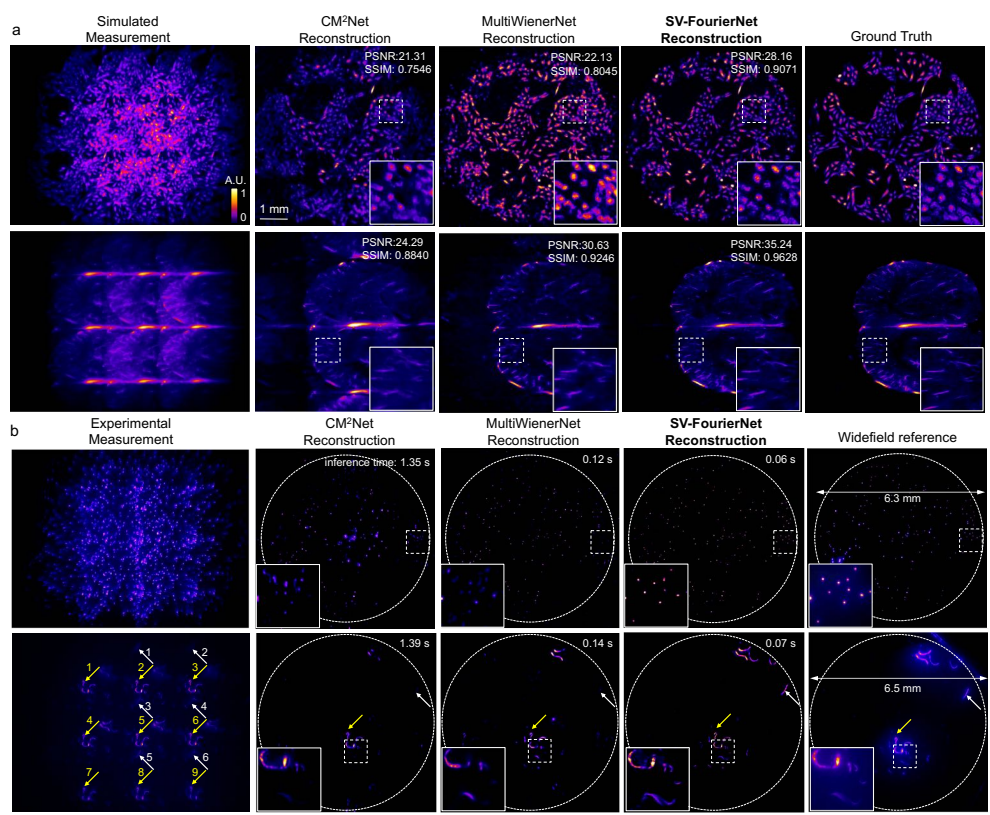

**Fig. S23. Comparison of SV-FourierNet with existing state-of-the-art networks on simulated and experimental datasets.** (a) Reconstruction results from simulated cell and vasculature measurements. SV-FourierNet achieves the best PSNR and SSIM for both types of samples. (b) Reconstruction results from experimental measurements on 10 $\mu$m fluorescent beads and freely moving *C. elegans*. SV-FourierNet provides the best visual quality and the fastest inference speed. Two *C. elegans* are indicated by the white and yellow arrows in both the measurement and the reconstructed images. The numbers on the measurement index the views from each microlens. SV-FourierNet successfully reconstructs the *C. elegans* populations across the entire FOV, while other networks suffer from loss of information especially near the FOV edges (marked by the white arrows).

28

## 8. ADDITIONAL RESULTS FOR RESOLUTION CHARACTERIZATION

### A. Resolution characterization of model-based algorithm and deep learning networks

This section compares the resolution of SV-FourierNet, MultiWienerNet and the LSV model-based algorithm by reconstructing the resolution target measurements captured at different FOV positions, as shown in Fig. S24. At the center of the FOV, all three deconvolution algorithms achieve a similar resolution of approximately 7.81 $\mu$m. Notably, SV-FourierNet is 2× and 60,000× faster than MultiWienerNet and the LSV model-based algorithm, respectively. At the edge of the FOV, the resolution for SV-FourierNet remains consistent, the resolution for LSV model-based algorithm degrades slightly, whereas the resolution of MultiWienerNet decreases significantly to 13.92 $\mu$m.

This study highlights that SV-FourierNet maintains a similar level of data fidelity and robustness across the FOV as the LSV model-based algorithm but with substantially faster deconvolution speed. Furthermore, when comparing SV-FourierNet with MultiWienerNet, it demonstrates significantly improved image quality at the FOV edge, underscoring the effectiveness of the learned global Fourier filters in managing shift-variant PSFs.

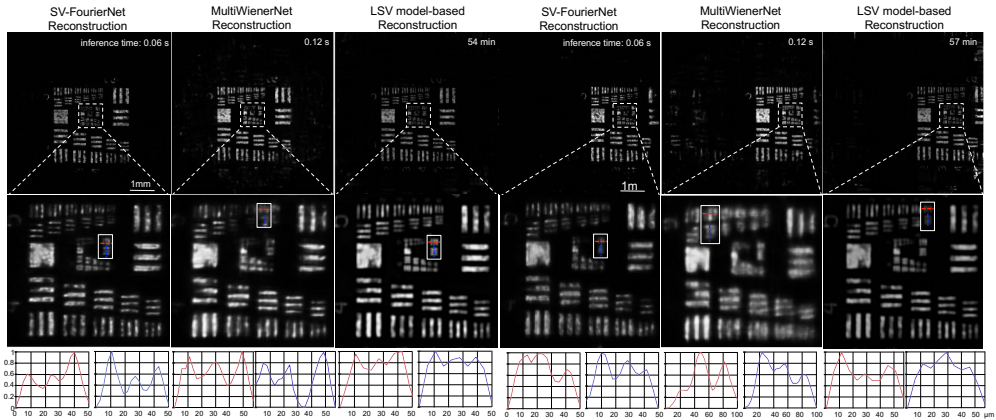

**Fig. S24. Comparison of resolution target reconstructions between SV-FourierNet, MultiWienerNet and LSV model based reconstruction.** SV-FourierNet, MultiWienerNet and LSV model-based algorithm achieve a resolution of 7.81 $\mu$m, 8.77 $\mu$m and 7.81 $\mu$m at the FOV center and resolution of 7.81 $\mu$m, 13.92 $\mu$m and 8.77 $\mu$m at the FOV edge.

### B. Resolution characterization at FOV edge over an EDoF

To demonstrate SV-FourierNet achieves consistent resolution within a 100 $\mu$m depth at the FOV edge, Fig. S25 presents reconstructions of a resolution target positioned at the FOV edge with an axial scanning step size of 10 $\mu$m. The reconstructions exhibit negligible resolution degradation within a 100 $\mu$m depth, corroborating the results shown in Fig. 4 of the manuscript. This demonstrates SV-FourierNet's capability for robust shift-variant deconvolution across a 100 $\mu$m depth range.

### C. Resolution characterization at different FOV positions

In this section, we characterize resolution across various FOV positions. Specifically, the resolution target is imaged at positions to the most right, left, center, down, and up of the FOV, shown in Fig. S26. SV-FourierNet consistently maintains a resolution of 7.8 $\mu$m across these varied FOV positions. However, a marginal contrast degradation is observed in comparison to the center of the FOV. This degradation can be attributed to pronounced shift variance arising from sensor truncation and shift variant aberrations, which current network configuration with 9 channels Fourier filters cannot entirely mitigate. To achieve a more uniform resolution distribution across the FOV, augmenting the number of channels could be explored, albeit with an associated increase in computational complexity, as discussed in Section 4C.

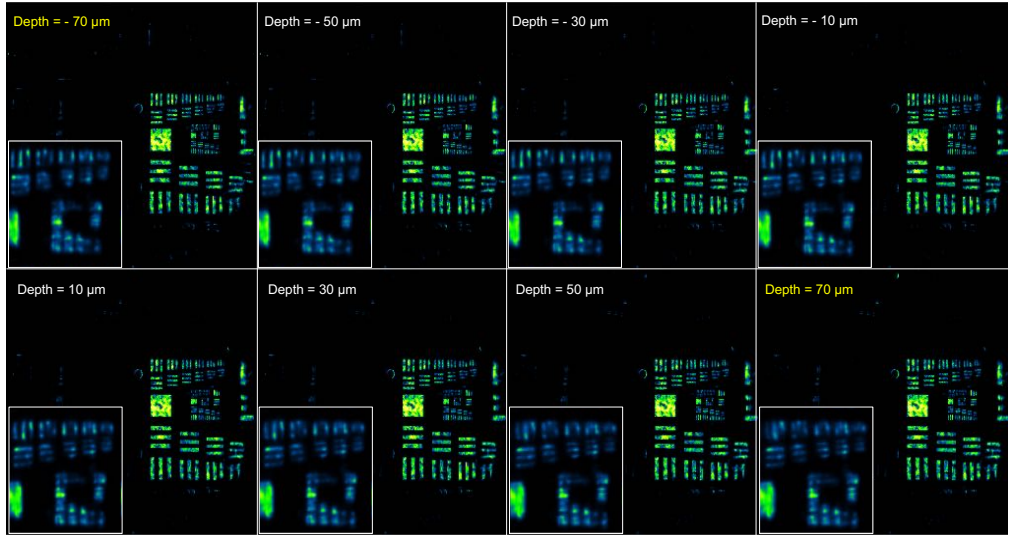

**Fig. S25. Resolution characterization at FOV edge across different depths.** SV-FourierNet consistently achieves 7.81 $\mu$m across a depth range of 100 $\mu$m.

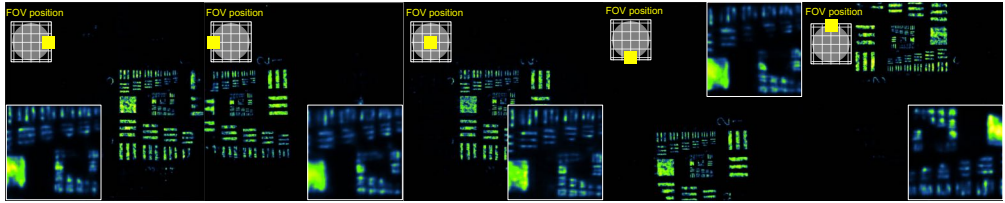

**Fig. S26. Resolution characterization across various FOV positions.** SV-FourierNet consistently achieves high reconstruction quality and comparable resolution across different positions within the FOV. The inset caption indicates the location of the resolution target within the FOV, with the yellow square denoting the resolution target and the shaded area indicating the FOV, marked by a grid with 1.5 mm interval.

### D. Resolution characterization with Power Spectral Density

This section presents power spectral density (PSD) analysis to evaluate the resolution consistency across the FOV and various experimental samples.

The resolution characterization across the FOV is demonstrated with reconstructions from simulated testing data. Each reconstruction is segmented into 64 patches (8x8 grid) with a patch size of 1.64 mm. Each patch represents a distinct FOV position. We record the center coordinates of these patches and calculate their radial displacement from the FOV center. The patches are grouped into four radial displacement bins [0 mm: 0.82 mm: 3.28 mm], and the averaged spectrum for each bin is computed. The PSD of these spectra, calculated on a logarithmic scale, is shown in Fig. S27(a). The PSD across different bins shows very similar profiles with slight energy degradation at higher resolution frequencies, indicating that SV-FourierNet maintains consistent resolution throughout the FOV. It is noted that patches with lower radial displacement exhibit higher noise levels due to fewer patches in those bins.

Furthermore, we assess the resolution consistency across different experimental samples to determine if SV-FourierNet exhibits bias towards specific sample types. Given that the energy levels of various samples can vary significantly, we normalize the energy of each reconstruction before computing the spectrum. The PSD results for different samples exhibit minimal variation, as illustrated in Fig. S27(b). The cut-off frequency, marked by the black dashed line, is at approximately 0.14 $\mu$m$^{-1}$ (7.14 $\mu$m), aligning with our resolution characterization on the resolution target. This comparison highlights the robust performance of SV-FourierNet across diverse experimental samples.

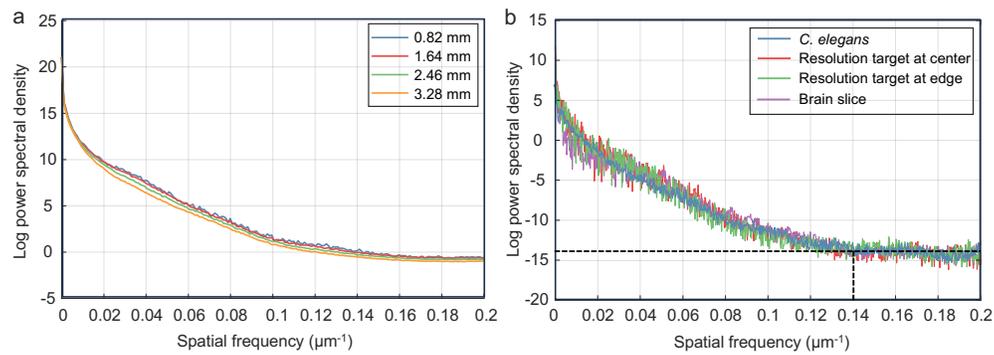

**Fig. S27. Resolution characterization via PSD.** (a) PSD analysis across various radial displacements are closely matched, demonstrating consistent resolution across the FOV. Patches with lower radial displacement exhibit higher noise due to fewer data points in those bins. (b) PSD analysis across different experimental samples exhibits minimal variation, indicating SV-FourierNet's robust performance across diverse experimental samples. The cut-off frequency, marked by a black dashed line is at approximately 0.14 $\mu m^{-1}$ (7.14 $\mu$m).

## 9. ADDITIONAL DETAILS ON EXPERIMENT

### A. Computational multi-aperture miniature setup

The multi-aperture miniscope is built with off-the-shelf and 3D-printed components, as shown in Fig. S28. The illumination module contains four symmetrically positioned, 45 degrees angled collimated LED units, each equipped with a lightweight 3D-printed free-form collimator suited between a surface-mounted LED (LXML-PB01-0040, Lumileds) and an excitation filter (no. 470, Chroma Technology). Collectively, the illumination module provides up to 1 mW/mm$^2$ of uniform illumination over an 8 × 8 mm$^2$ FOV. The prototype uses an off-the-self plastic 3×3 MLA (no. 630, Fresnel Technologies Inc.), forming a finite conjugate imaging geometry with a working distance of ∼7.5 mm and an MLA-sensor distance of ∼4.5 mm. Given the working distance and the MLA pitch size, the effective numerical aperture (NA) for a single lens is ∼0.05, giving an optical resolution of 5.4 $\mu$m. For green fluorescence imaging, a hybrid filter set is incorporated, combining an interference filter (no. 535/50, Chroma Technology) placed in front of the MLA and a long-pass absorption filter (Wratten color filter no. 12, Edmund Optics) placed after the MLA. It uses a backside-illuminated (BSI) CMOS sensor (IM226, IDS Imaging) with 12 million pixels and 1.85 $\mu$m pixel size, achieving an effective pixel size of 2.8 $\mu$m and a Nyquist sampling resolution of 5.6 $\mu$m.

Housed within a 3D-printed enclosure (printed with Formlabs Form 2, black resin, no. RS-F2GPBK-04), the prototype features size of ∼30 × 30 × 15 mm$^3$ and a total weight of ∼9.5 g, primarily due to the CMOS sensor. The custom parts excluding the PCB are ∼20 × 20 × 13 mm$^3$ in size and weigh only ∼2.5 g, showcasing its portability and compactness for diverse imaging applications.

### B. Wide-field reference system setup

To validate the reconstruction results for freely moving *C. elegans*, we set up a widefield fluorescence microscope to record fluorescence simultaneously with our miniscope, as shown in Fig. S28. This widefield setup includes a 4F system (comprising Thorlabs AC508-180-A and Thorlabs ACT508-250-A) and utilizes the same type of sensor as the miniscope. Both sensors were synchronized to capture fluorescence signals.

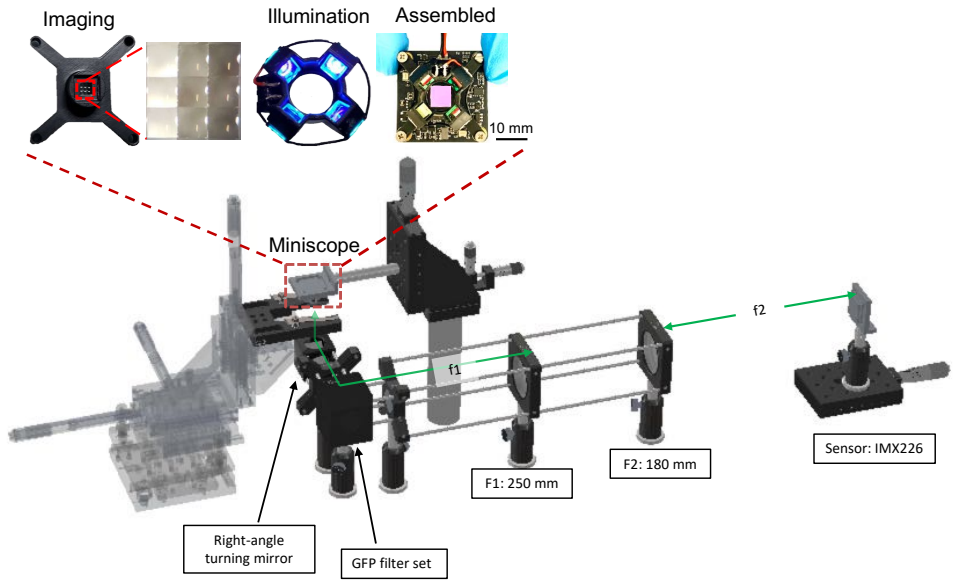

**Fig. S28.** Experimental set up for computational multi-aperture miniscope and wide field reference

### C. Preparation of brain slice

Ex-vivo fixed rodent brain slices with GFP-expressing neurons in the bed nucleus of the stria terminalis (BNST) were prepared with viral-mediated techniques on male C57Bl/6 mice. Following

anesthesia and preoperative analgesia, a small craniotomy was performed bilaterally over each injection site. Viral solutions (100-200 nL, pAAV-CAG-GFP, Addgene # 37825-AAVrg) were delivered to the BNST and labeled both local neurons at the injection site in BNST as well as upstream brain areas. Post-recovery, the animals were anesthetized and perfused with paraformaldehyde, and their brains were sectioned coronally at 75 μm thickness. The sections were then mounted for analysis.

This study was performed in strict accordance with the recommendations in the Guide for the Care and Use of Laboratory Animals of the National Institutes of Health. All animals were handled according to approved Institutional Animal Care and Use Committee (IACUC) protocols (#201800540) of Boston University.

### D. Preparation of *C. elegans*

The transgenic Caenorhabditis elegans (*C. elegans*) strain CL2355 [pCL45 (snb-1::Abeta 1-42::3' UTR(long) + mtl-2::GFP], genetically modified to express green fluorescent protein (GFP), are grown on nematode growth medium (NGM) agarose plates. These plates are seeded with Escherichia coli K-12 (Carolina Biological Supply #124500) to ensure a stable food supply. For imaging, a small portion of the *C. elegans* are transferred into an ∼8 mm × 8 mm section formed by 2% agarose gel (Sigma-Aldrich) on a glass microscope slide. A 20 μL drop of S-Basal solution is then added to this section to enable the *C. elegans* to move freely within the solution. After preparation, the slide is immediately placed on the microscope stage for imaging.

### E. Additional result of brain slice section

To provide a clearer visual comparison between the reconstruction and the ground truth of the brain slice section, we overlay the reconstruction (depicted in red) onto the ground truth (depicted in green), which was captured using a 2× 0.1NA objective. In areas where the reconstruction aligns accurately with the ground truth, the overlapping regions appear yellow. We have zoomed into four regions on both sides of the brain section, all demonstrating a high degree of correspondence. This observation further substantiates the robustness of SV-FourierNet in reconstructing complex brain tissues.

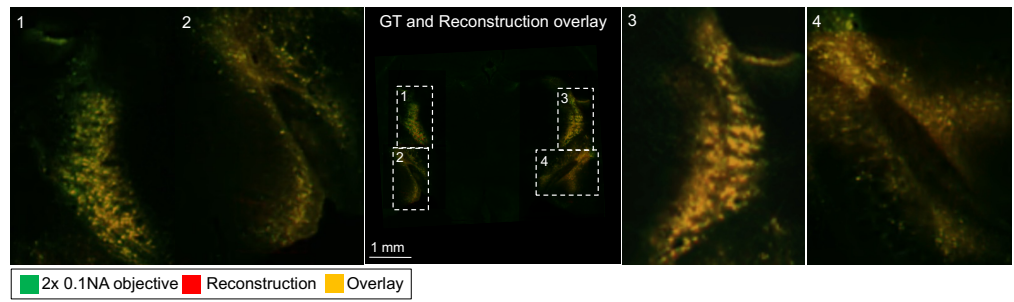

**Fig. S29. Overlay of SV-FourierNet reconstruction and ground truth measurement from wide field 2× 0.1 NA objective on brain slice section.**

33


\end{document}